\documentclass[prc,aps,superscriptaddress,twocolumn,showpacs,nofootinbib]{revtex4}

\usepackage{graphicx,amsmath,amssymb,bm,multirow}
\usepackage{amsthm,mathrsfs}
\usepackage{amsfonts}
\usepackage{dcolumn}
\usepackage{bm}
\usepackage{float}
\usepackage[colorlinks,linkcolor=blue,anchorcolor=blue,citecolor=blue]{hyperref}
\usepackage[usenames,dvipsnames]{color}

\newcommand{\balp}{{\boldsymbol{\alpha}}}  
\newcommand{\bnab}{{\mathbf{\nabla}}}
\newcommand{\br}{\mbox{\boldmath $r$}}

\newcommand{\bp}{{\mathbf{p}}}
\newcommand{\bP}{{\mathbf{P}}}

\newcommand{\beq}{\vspace{0.5em}\begin{equation}}
\newcommand{\eeq}{\end{equation}\vspace{0.5em}}
\newcommand{\beqn}{\vspace{0.5em}\begin{eqnarray}}
\newcommand{\eeqn}{\end{eqnarray}\par\vspace{0.5em}\noindent}

\newcommand{\bsub}{\begin{subequations}}
\newcommand{\esub}{\end{subequations}}

\begin{document}
 \preprint{preprint}

\title{Triaxially deformed relativistic point-coupling model for $\Lambda$ hypernuclei: a quantitative analysis of hyperon impurity effect on nuclear collective properties}
\author{W. X. Xue}
\affiliation{School of Physical Science and Technology, Southwest University, Chongqing 400715, China}
\author{J. M. Yao} \email{jmyao@swu.edu.cn}
\affiliation{School of Physical Science and Technology, Southwest University, Chongqing 400715, China}
\affiliation{Department of Physics, Tohoku University, Sendai 980-8578, Japan}
\author{K. Hagino}
\affiliation{Department of Physics, Tohoku University, Sendai 980-8578, Japan}
\affiliation{Research Center for Electron Photon Science, Tohoku University, 1-2-1 Mikamine, Sendai 982-0826, Japan}
\author{Z. P. Li}
\affiliation{School of Physical Science and Technology, Southwest University, Chongqing 400715, China}
\author{H. Mei}
\affiliation{Department of Physics, Tohoku University, Sendai 980-8578, Japan}
\affiliation{School of Physical Science and Technology, Southwest University, Chongqing 400715, China}
\author{Y. Tanimura}
\affiliation{Institut de Physique Nucl\'eaire, IN2P3-CNRS, Universit\'e Paris-Sud, F-91406 Orsay Cedex, France}

\begin{abstract}
\begin{description}

\item[Background] The impurity effect of hyperon on atomic nuclei has
received a renewed interest in nuclear physics since the first
experimental observation of appreciable
reduction of $E2$ transition strength in low-lying states of
hypernucleus $^{7}_\Lambda$Li. Many more data on low-lying states of $\Lambda$ hypernuclei
will be measured soon for $sd$-shell nuclei, providing good opportunities to study the $\Lambda$
impurity effect on nuclear low-energy excitations.

\item[Purpose] We carry out a quantitative analysis of $\Lambda$ hyperon impurity effect on
the low-lying states of $sd$-shell nuclei at the beyond-mean-field level
based on a relativistic point-coupling energy
density functional (EDF), considering that
the $\Lambda$ hyperon is injected into the lowest positive-parity
($\Lambda_s$) and negative-parity ($\Lambda_p$) states.

\item[Method]  We adopt a triaxially deformed relativistic mean-field (RMF) approach for hypernuclei
and calculate the $\Lambda$ binding energies of hypernuclei as well as
the potential energy surfaces (PESs) in $(\beta, \gamma)$ deformation plane.
We also calculate the PESs for the $\Lambda$ hypernuclei with good quantum numbers
using a microscopic particle rotor model (PRM) with
the same relativistic EDF. The triaxially deformed RMF approach is further
applied in order to determine the parameters of a five-dimensional collective Hamiltonian (5DCH)
for the collective excitations of triaxially deformed core nuclei.
Taking $^{25,27}_{~~~~\Lambda}$Mg and $^{31}_{~\Lambda}$Si as examples,
we analyse the impurity effects of $\Lambda_s$ and $\Lambda_p$ on the low-lying states of the core nuclei.

\item[Results]
We show that $\Lambda_s$ increases the excitation energy of the $2^+_1$ state and decreases the $E2$
transition strength from this state to the ground state by $12\%-17\%$.
On the other hand, $\Lambda_p$
tends to develop pronounced energy minima with larger deformation, although
it modifies the collective parameters in such a way that the collectivity of the core
nucleus can be either increased or decreased.

\item[Conclusions] The quadrupole deformation significantly affects the $\Lambda$
binding energies of deformed hypernuclei.
A beyond-mean-field approach with the dynamical correlations due to restoration of broken symmetries
and shape fluctuation is essential in order to study the $\Lambda$ impurity effect in a quantitative way.

\end{description}

\end{abstract}

\maketitle

 \section{\label{introduction}Introduction}
Since the first discovery of $\Lambda$ hypernuclei in 1953
\cite{Danysz1953_PhilosMag44-348,Danysz1953_Bull.Acad.Pol},
hypernuclear physics as an important branch of nuclear physics has attracted
lots of attention and many novel phenomena have been discovered in this field.
Due to the absence of Pauli exclusion principle from other nucleons,
a hyperon with strangeness degree-of-freedom can probe deeply into the interior of nuclear medium
and significantly modify nuclear properties.
For example, a hyperon may induce
a softening of the equation of state of nuclear matter changing the properties of neutron
stars~\cite{Glendenning2000_NY}, a shrinkage of the size of atomic nuclei with cluster
structure~\cite{MBI83,HKMM99,Tanida2001_PRL86-1982}, a stabilization of the binding of unbound nuclear
systems~\cite{HKMYY96} and thus the driplines of nucleons~\cite{VPLR98},
a modification of nuclear deformation~\cite{WH08,Lu2011_PRC84-014328} and
collective excitations~\cite{Yao2011_NPA868-12,Isaka12,MHYM14,HYM13}, and a reduction of
fission-barrier height in heavy nuclei~\cite{MCH09,MC11}.
Because hyperon-nucleon and hyperon-hyperon scattering experiments are difficult to perform,
the study of properties of hypernuclei has in fact been playing a vital role in understanding baryon-baryon
interactions in nuclear medium, which are important not only for understanding hypernuclear structure
but also for the study of hypernuclear matter and neutron stars~\cite{Hofmann2001_PRC64-025804}.
A comprehensive introduction to the history and/or recent developments on various aspects
in hypernuclear physics can be found in the review
papers~\cite{Davis2005,Dalitz2005,Hashimoto2006_PPNP-564,Hiyama2009PPNP,Schulze2010NPA,Hiyama2012FBS,Botta2012EPJA,Hagino2014}.

Thanks to the advent of hyperball facility for measuring hypernuclear $\gamma$-ray spectroscopy
with high resolution~\cite{Tamura2000PhysRevLett.84.5963,Tanida2001_PRL86-1982},
the study of $\Lambda$ hyperon impurity effect on nuclear deformation and low-energy structure
has attracted a renewed interest. The self-consistent mean-field (SCMF) approaches make good tools for
this study because they provide a vivid way to investigate how the ground-state deformation
is affected by adding a $\Lambda$ particle.  In the past decades, the SCMF approaches have
been adopted extensively to study the structure of hypernuclei~\cite{Rayet76,Brockmann1977_PLB69-167,
Boguta1981,YBZ88,MZ89,Rufa1990PhysRevC.42.2469,VPLR98,ST94,MJ94,Keil2000PhysRevC.61.064309,Lu2002CPL,
Finelli2009NPA,Song2010PK1-Y1}. However, most of these studies are restricted to spherical systems.
In recent years, the SCMF approaches have been extended to deformed cases in order to examine the
change of nuclear deformation after adding a hyperon, based either on Skyrme
forces~\cite{Zhou2007_PRC76-034312,WHK11,Li2013_PRC87-014333} or on effective relativistic meson-exchange
Lagrangian~\cite{WH08,Lu2011_PRC84-014328,Lu2014PhysRevC.89.044307,Xu2015}. It has been
found in these mean-field studies that the deformations of hypernuclei and the corresponding core nucleus
are rather similar, but with some exceptions as predicted by the relativistic mean-field (RMF)
calculations~\cite{WH08,Lu2011_PRC84-014328}.
It implies that the hyperon impurity effect is generally larger in the calculation
with relativistic approaches than that with non-relativistic approaches, as has been pointed
out by Schulze {\it et al.}~\cite{Schulze2010PTP}.
Therefore, one would encounter more
opportunities to see drastic deformation changes from ordinary nuclei to hypernuclei in
the studies based on relativistic energy density functionals (EDFs).

It should be pointed out that, in most of the previous SCMF studies allowing deformation,
the $\Lambda$ hyperon is put in the lowest positive-parity ($\Lambda_s$) state. The corresponding
$\Lambda_s$-hypernuclei turn out to have a softer energy surface than their core nucleus.
This indicates that the dynamical shape fluctuation effect will be more important in hypernuclei than
in normal nuclei, and thus the mean-field approaches might overestimate/underestimate the $\Lambda$ hyperon
impurity on nuclear deformation and shapes. To quantify the $\Lambda$ hyperon impurity effect, one therefore has to go beyond-mean-field (BMF) approximation
to take into account the dynamical correlation effects associated with symmetry restoration and shape fluctuation.
Notice that the deformed mean-field breaks rotational symmetry, and,
if one works only at the mean-field level, the connection of the deformed solution
to spectroscopic observables, such as $B(E2)$ value, has to rely on additional assumptions
such as the rigid-rotor model, which becomes ill-defined in light and soft nuclei.

Recently, we have quantitatively studied the $\Lambda$ impurity effect on the low-lying states
of $^{24}$Mg by using a five-dimensional collective Hamiltonian (5DCH) as a choice of the BMF
approaches~\cite{Yao2011_NPA868-12}. The parameters of the 5DCH were determined by a triaxially
deformed Skyrme-Hartree-Fock (SHF)+BCS calculation. We have found that the presence of one
$\Lambda$ hyperon in the lowest positive-parity state reduces the $B(E2:2^+_1\to 0^+_1)$ in $^{24}$Mg
by 9\% and shifts up the excitation energy of the second $2^+$ state by about 240 keV.
Similar conclusions have also been found in the BMF study based on the antisymmetrized
molecular dynamics (AMD) model~\cite{IKDO11,Isaka12}.
We note that these BMF studies are within non-relativistic frameworks.
Moreover, the impurity effect of $\Lambda$ hyperon in the lowest negative-parity ($\Lambda_p$)
state has not been well examined in the 5DCH approach.

In view of the above facts, it is interesting to quantitatively study the $\Lambda$ hyperon impurity effect
based on a relativistic EDF at the BMF level by putting the $\Lambda$ hyperon in the $\Lambda_s$
and $\Lambda_p$ states. To this end, as a continuation of our previous work~\cite{Yao2011_NPA868-12},
we adopt the same 5DCH approach for the low-lying states of core nuclei but with collective parameters determined
from a triaxial RMF+BCS calculation. We generalize our triaxial RMF approach in a three-dimensional
harmonic-oscillator (3DHO) basis for ordinary nuclei~\cite{Yao2009PhysRevC.79.044312}
to $\Lambda$ hypernuclei by including $\Lambda$ hyperons.
This 5DCH method based on the triaxial RMF solutions can be regarded as the Gaussian overlap approximation to
the generator coordinate method (GCM)~\cite{Yao10,Yao11} with three-dimensional angular momentum
projection (3DAMP)~\cite{Yao09}. The success of the 5DCH method based on relativistic EDFs has been illustrated in a series of
calculations for spherical, transitional, and deformed nuclei from light to
superheavy regions~\cite{Li2009PhysRevC.79.054301,Li2010PhysRevC.81.064321,Li2011PhysRevC.84.054304,
Mei2012PhysRevC.85.034321,Fu2013PhysRevC.87.054305}.
In particular, the validity of the 5DCH approach for the low-lying states of $^{76}$Kr has recently been
verified against a seven-dimensional GCM calculation~\cite{Yao14}.

It is worth mentioning that there are two other triaxially deformed RMF codes
for $\Lambda$ hypernuclei based on meson-exchange interaction.
One was developed by B. N. L\"{u} {\em et al.} with an axially deformed HO
basis~\cite{Lu2011_PRC84-014328,Lu2014PhysRevC.89.044307}, while the other one was developed by
H. F. L\"{u} {\em et al.}~\cite{Sang2013.PhysRevC.88.064304} based on the triaxial RMF code with
the 3DHO basis for ordinary nuclei~\cite{Yao06} with time-odd fields but without pairing correlation.
Our triaxial code developed in the present work includes the paring correlation and
is mainly based on but not restricted to the relativistic
point-coupling EDFs, which have been widely adopted to study nuclear low-lying
states within the framework of multi-reference covariant density functional theory
(CDFT)~\cite{Yao10,Yao14,Song14,Yao14DBD}.

The paper is arranged as follows. In Sec.~\ref{Sec.II}, we present the main formalism of the triaxial
RMF approach for $\Lambda$ hypernuclei. In Sec.~\ref{Sec.III}, we present the results for $\Lambda$ binding
energies obtained with the triaxial RMF code and compare to the results of the spherical code.
In particular, we discuss the potential energy surfaces (PESs) for $\Lambda$ hypernuclei $^{25,27}_{~~~~\Lambda}$Mg, $^{31}_{~\Lambda}$Si as well as the core nucleus of each of these hypernuclei in $(\beta, \gamma)$ deformation plane.
In Sec.~\ref{Sec.IV}, the microscopic particle-rotor model (PRM)  for the PES of
$\Lambda$ hypernucleus $^{25}_{~\Lambda}$Mg with spin-parity of $I^\pi=1/2^+$ and $1/2^-$ are discussed
in comparison with that of $^{24}$Mg with $J^\pi=0^+$. In Sec.~\ref{Sec.V}, the 5DCH method is adopted to study a change in the low-lying states of the core nucleus by adding a $\Lambda$ hyperon. The impurity effect of $\Lambda$ hyperon is discussed both for
$\Lambda_s$ and $\Lambda_p$. A summary of the present study and an outlook are then given in Sec.~\ref{Sec.VI}.

 \section{Triaxially deformed relativistic mean-field approach for $\Lambda$ hypernuclei}
 \label{Sec.II}

 In the present triaxial RMF approach for $\Lambda$ hypernuclei,
 the nucleon-nucleon ($NN$) and nucleon-hyperon ($N\Lambda$)
 effective interactions are described in terms of contact couplings
 with different vertices. The Lagrangian density for $\Lambda$ hypernuclei
 then reads,

\begin{eqnarray}
\label{Lagrangian}
{\cal L} ={\cal L}^{\rm free} + {\cal L}^{\rm em} + {\cal L}^{NN} + {\cal L}^{N\Lambda}+ {\cal L}^{\Lambda\Lambda},
\end{eqnarray}
where the first term ${\cal L}^{\rm free}$ is for the free nucleons and
hyperon, and ${\cal L}^{\rm em}$ for the Coulomb interaction between protons.
The third term ${\cal L}^{NN}$ is for the $NN$ effective interaction part.
The detailed expressions for these terms can be found for example in
Refs.~\cite{Burvenich2002_PRC65-044308,Zhao2010PhysRevC.82.054319}.
Since we focuse on single-$\Lambda$ hypernuclei in this work,
the ${\cal L}^{\Lambda\Lambda}$ term for $\Lambda\Lambda$ interaction
vanishes. The $N\Lambda$ interaction is chosen as in
Ref.~\cite{Tanimura2012_PRC85-014306}, that is,
\begin{eqnarray}
\begin{aligned}
{\cal L}^{N\! \Lambda}=
{\cal L}_{\rm 4f}^{N\! \Lambda}+{\cal L}_{\rm der}^{N\!\Lambda}
+{\cal L}_{\rm ten}^{N\!\Lambda},
\end{aligned}
\end{eqnarray}
with
 \bsub
 \label{LO}
\begin{eqnarray}
{\cal L}_{\rm 4f}^{N\!\Lambda}
&=&
-\alpha_S^{(N\!\Lambda)}(\bar{\psi}^{N}\psi^N)(\bar{\psi}^{\Lambda}\psi^{\Lambda})\nonumber\\
&&
-\alpha_V^{(N\!\Lambda)}(\bar{\psi}^{N}\gamma_{\mu}\psi^N)
(\bar{\psi}^{\Lambda}\gamma^{\mu}\psi^{\Lambda}), \\
{\cal L}_{\rm der}^{N\!\Lambda}
&=&-\delta_S^{(N\!\Lambda)}(\partial_{\mu}\bar{\psi}^{N}\psi^N)
(\partial^{\mu}\bar{\psi}^{\Lambda}\psi^{\Lambda}) \nonumber\\
&& -\delta_V^{(N\!\Lambda)}(\partial_{\mu}\bar{\psi}^{N}\gamma_{\nu}\psi^N)
(\partial^{\mu}\bar{\psi}^{\Lambda}\gamma^{\nu}\psi^{\Lambda}),\\
\label{NL}
{\cal L}_{\rm ten}^{N\!\Lambda}
&=& -\alpha^{(N\!\Lambda)}_T(\bar{\psi}^{\Lambda}\sigma^{\mu\nu}\psi^{\Lambda})
(\partial_{\mu}\bar{\psi}^{N}\gamma_{\nu}\psi^N).
\end{eqnarray}
\esub
We note that the vector-meson-like tensor coupling term,
${\cal L}_{\rm ten}^{N\!\Lambda}$, is usually adopted to reproduce the
smallness of spin-orbit splitting in $\Lambda$ single-particle
spectra~\cite{Noble80,Boussy81,ZQ85,Jennings1990Phys.Lett.B246.325,ST94,Yao2008,Tanimura2012_PRC85-014306},
although it can also be explained in terms of an almost complete
cancelation between short-range scalar and vector contributions and
longer range terms generated by two-pion exchange~\cite{Finelli2009NPA}.
The Lagrangian contains sixteen coupling constants
$\alpha_S$, $\alpha_V$, $\alpha_{TS}$, $\alpha_{TV}$, $\alpha_S^{(N\Lambda)}$,
$\alpha_V^{(N\Lambda)}$, $\alpha_T^{(N\Lambda)}$, $\beta_S$, $\gamma_S$,
$\gamma_V$, $\delta_S$, $\delta_V$, $\delta_{TS}$, $\delta_{TV}$,
$\delta_S^{(N\Lambda)}$ and $\delta_V^{(N\Lambda)}$, which are usually
optimized at the mean-field level to properties of several atomic
nuclei and hypernuclei.

From the Lagrangian density (\ref{Lagrangian}), one can derive the
corresponding energy $E_{\textrm{RMF}}$ at the mean-field level with the {\em no-sea} approximation, which can
be decomposed into two parts: the pure nuclear part and the $\Lambda$ hyperon part,
 \beq
 \label{EDF}
 E_{\textrm{RMF}} = E^{N}_{\textrm{RMF}} + E^{\Lambda}_{\textrm{RMF}},
 \eeq
with
\bsub\begin{eqnarray}
E^{N}_{\textrm{RMF}}  &=& T_N + \int d^3r\left[\varepsilon_{NN}(\textbf{r})+ \frac{1}{2}A_0 e\rho_V^{(p)}\right], \\
E^{\Lambda}_{\textrm{RMF}}
&=& T_\Lambda +\int d^3r\varepsilon_{N\Lambda}(\textbf{r}).
\end{eqnarray}\esub
The first term in these equations,
$T_{B}={\rm Tr}[ (\balp \cdot\bp +\gamma^0m_B)\rho^B_V]$,
is for the kinetic energy of nucleons ($B=N$) or hyperon ($B=\Lambda$),
where $m_B$ is the corresponding mass.  For the sake of simplicity,
time-reversal invariance is usually imposed in the mean-field
calculations for the $\Lambda$ hypernuclei, in which case,
the $NN$ and $N\Lambda$ interaction terms are given by,
 \bsub\begin{eqnarray}
\label{Energy:NN}
\varepsilon_{NN}
&=&\frac{1}{2}\sum\limits_{m=S,V,TS,TV} \left[\alpha_m(\rho_m^N)^2
  + \delta_m\rho_m\Delta\rho_m^N\right] \nonumber \\
&&+\frac{1}{3}\beta_S(\rho_S^N)^3+\frac{1}{4}\gamma_S(\rho_S^N)^4 +\frac{1}{4}\gamma_V(\rho_V^N)^4, \\
\label{Energy:NL}
 \varepsilon_{N\Lambda}
 &=& \sum\limits_{m=S,V}\left[\alpha_m^{(N\Lambda)}\rho_m^N\rho_m^\Lambda+\delta_m^{(N\Lambda)}\rho_m^N\Delta\rho_m^\Lambda\right] \nonumber \\
 &&+\alpha_T^{(N\Lambda)}\rho_V^N\rho_T^\Lambda,
\end{eqnarray}\esub
respectively.
In these equations, the densities $\rho^B_{m}$ and the tensor density $\rho^\Lambda_T$
are defined as
\begin{eqnarray}
\label{rhotensor}
\rho^B_m=\sum \limits_kv^2_k\bar\psi^B_k\Gamma_m \psi^B_k, ~~
\rho^\Lambda_T= \bnab\cdot (\bar\psi^\Lambda  i \balp \psi^\Lambda),
\end{eqnarray}
where the vertex $\Gamma_m$ is $1, \gamma^0, \tau_3$, and $\gamma^0\tau_3$,
with the index $m$ running over $S, V$, $TS$ and $TV$,
which represents respectively the isoscalar-scalar, isoscalar-vector,
isovector-scalar and  isovector-vector types of coupling characterized
by their transformation properties in isospin-Lorentz spaces.
$v^2_k$ is the occupation probability of the $k$-th single-particle
energy level of neutrons or protons to be determined by the BCS method.
The $\balp$ and $\gamma^\mu$ are the $4\times4$ Dirac matrices.

Minimization of the RMF energy (\ref{EDF}) with respect to the
single-particle wave function for nucleons or hyperon leads to the
Dirac equation, which reads,
\begin{eqnarray}
\label{Dirac:N}
\left[\balp\cdot\bp+V^N_0 +\gamma^0(m_N+S^N)\right]\psi^N_k(\br)=\epsilon_k^N\psi^N_k(\br),
\end{eqnarray}
for nucleons, with the scalar field  $S^N(\br)=\Sigma_S(\br)+\tau_3 \Sigma_{TS}(\br)$ and vector field $V^N_0(\br)=\Sigma_V(\br)+\tau_3\Sigma_{TV}(\br)$ defined as
 \bsub\begin{eqnarray}\label{dirac-nucleon1}
\Sigma_S&=&\alpha_S\rho_S^N+\beta_S(\rho_S^N)^2+\gamma_S(\rho_S^N)^3+\delta_S\Delta\rho_S^N\\
&&+\alpha_S^{(N\Lambda)}\rho_S^\Lambda+\delta_S^{(N\Lambda)}\Delta\rho_S^\Lambda, \\
\label{dirac-nucleon2}
\Sigma_{TS}&=&\delta_{TS}\Delta\rho_{TS}^N+\alpha_{TS}\rho_{TS}^N, \\
\label{dirac-nucleon3}
\Sigma_V&=&\alpha_V\rho_V^N+\gamma_V(\rho_V^N)^3+\delta_V\Delta\rho_V^N
+eA_0\frac{1-\tau_3}{2}\\
&&+\alpha_V^{(N\Lambda)}\rho_V^\Lambda+\delta_V^{(N\Lambda)}\Delta\rho_V^\Lambda+\alpha_T^{(N\Lambda)}\rho_T^\Lambda, \\
\label{dirac-nucleon4}
\Sigma_{TV}&=&\alpha_{TV}\rho_{TV}^N+\delta_{TV}\Delta\rho_{TV}^N.
\end{eqnarray}
\esub
On the other hand,
the Dirac equation for the $\Lambda$ hyperon inside the hypernucleus reads
 \begin{eqnarray}
 \label{Dirac:L}
\left[\balp\cdot\bp+V^\Lambda_0+\gamma^0(S^\Lambda+m_\Lambda)\right]\psi^\Lambda_k(\br) =\epsilon_k^\Lambda\psi^\Lambda_k(\br)
\end{eqnarray}
with the vector field $V^\Lambda_0=U_V+ U_T$, and
\bsub\begin{eqnarray}\label{dirac-Lambda}
U_V&=&\delta_V^{(N\Lambda)}\Delta\rho_V^N+\alpha_V^{(N\Lambda)}\rho_V^N,\\
\label{U_T}
U_T&=&-i\alpha_T^{(N\Lambda)}\beta\balp\cdot \bnab\rho_V^N,\\
S^\Lambda &=&\delta_S^{(N\Lambda)}\Delta\rho_S^N+\alpha_S^{(N\Lambda)}\rho_S^N.
\end{eqnarray}
\esub

These two Dirac equations, Eqs. (\ref{Dirac:N}) and (\ref{Dirac:L}),
are solved by expanding the Dirac spinors $\psi^B_k$ for nucleons
and hyperon on the basis of a 3DHO with the oscillator length parameter
chosen as $b_x=b_y=b_z=\sqrt{\hbar/m\omega_0}$, where the oscillator
frequency is determined by $\hbar\omega_0=41A^{-1/3}$ (MeV). In addition,
to reduce the computational task, it is assumed that the total densities
are symmetric under reflections with respect to the three planes
$xy$, $xz$ and $yz$.  The Coulomb field $A_0$ is obtained through a direct
integration of the Poisson equation.  To obtain the total energies
and the mean-field wave functions for triaxially deformed hypernuclei
and the corresponding core nucleus as a function of deformation
parameters ($\beta, \gamma$), a quadratic constraint calculation
on the mass quadrupole moments is carried out by minimizing the
following energy with respect to single-particle wave function,
\begin{equation}
E'=E_{\rm RMF} +\sum_{\mu=0,2}C_{2\mu}(\langle \hat{Q}_{2\mu}\rangle - q_{2\mu})^2,
\label{constraint}
\end{equation}
where $C_{2\mu}$ is a stiffness parameter and $\langle \hat{Q}_{2\mu}\rangle$ denotes the expectation value of the mass quadrupole moment operator,
\begin{equation}
 \hat{Q}_{20}  =  \sqrt{\dfrac{5}{16\pi}} (2z^2 - x^2 - y^2),~~
  \hat{Q}_{22}  = \sqrt{\dfrac{15}{32\pi}} (x^2 - y^2).
\end{equation}
In Eq. (\ref{constraint}),
$q_{2\mu}$ are  the quadrupole moment of mean-field state to be obtained.
The deformation parameters ($\beta, \gamma$) of mean-field state
are related to the expectation values of the mass quadrupole moment
operator by
$\beta= \dfrac{4\pi}{3AR^2}\sqrt{\langle Q_{20}\rangle^2+2\langle Q_{22}\rangle^2}$
and $\gamma=\tan^{-1}\left(\sqrt{2}\,\dfrac{Q_{22}}{Q_{20}}\right)$,
respectively, with $R=1.2A^{1/3}$ (fm). We note that the deformation
parameters ($\beta, \gamma$) are calculated with the nuclear
density $\rho^N(\br)$ for the core nuclei (cn) and with the total
density $\rho^N(\br)$+$\rho^\Lambda(\br)$ for the hypernuclei.

The center-of-mass correction energy $E_{\rm cm}$ is calculated by taking
the expectation value of the kinetic energy for the center-of-mass motion
with respect to the mean-field wave function. For a single-$\Lambda$
hypernucleus, it is given by
\begin{equation}
  E_{\rm cm}=\frac{ \langle \bP^2_N + \bP^2_\Lambda\rangle}{2(Am_N+m_{\Lambda})},
\label{cmenergy}
\end{equation}
where $\bP_B$ is the total momentum of the baryons ($B=N, \Lambda$)
in hypernucleus with $A$ nucleons and one $\Lambda$ hyperon.

Following Refs.~\cite{Burvenich2002_PRC65-044308,Yao2009PhysRevC.79.044312},
the pairing correlation among nucleons is taken into account with the
BCS method using a density-independent zero-range pairing force
supplemented with a smooth cutoff~\cite{Bender00}.
The resultant pairing energy $E_{\rm pair}$ is added to the total energy,
which is finally given by
 \beq
   E_{\rm tot} (^{A+1}_{~~~\Lambda} Z)=  E_{\rm RMF} - Am_Nc^2 - m_\Lambda c^2 - E_{\rm cm} + E_{\rm pair}.
 \eeq
 The total binding energy $B (^{A+1}_{~~~\Lambda} Z)$ of a
 single-$\Lambda$ hypernucleus is given by
 $B(^{A+1}_{~~~\Lambda} Z) = -  E_{\rm tot} (^{A+1}_{~~~\Lambda} Z)$.
 To study the change of energy surface of nuclear core by the
 $\Lambda$ hyperon, we also introduce the energy
 $E^{\rm cn}_{\rm tot} (^{A+1}_{~~~\Lambda} Z)$ for the core nucleus inside
a hypernucleus as
  \beq
  E^{\rm cn}_{\rm tot} (^{A+1}_{~~~\Lambda} Z)
  \equiv E_{\rm tot} (^{A+1}_{~~~\Lambda} Z) - E^\Lambda_{\rm RMF} + E^\Lambda_{\rm cm},
  \eeq
  where the last term,
  $E^\Lambda_{\rm cm}=\dfrac{\langle\bP^2_\Lambda\rangle}{2(Am_N+m_{\Lambda})}$,
  is the contribution of $\Lambda$ particle to the center-of-mass correction energy.

\begin{figure}[]
\centering
\includegraphics[width=9cm]{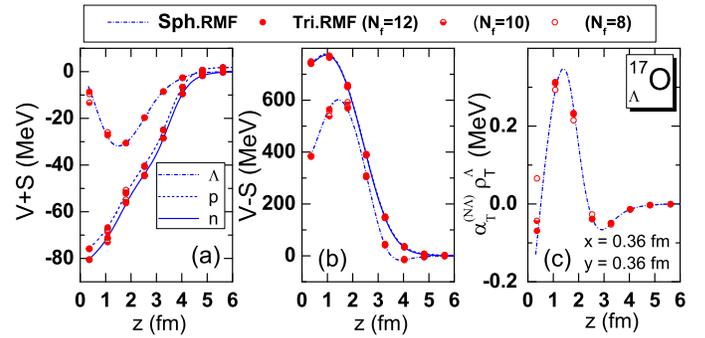}
\caption{(Color online) The convergence feature of
  potentials $V^B_0 (\br)+S^B(\br)$ (a), $V^B_0(\br)-S^B(\br)$ (b)
  (with $B=n, p, \Lambda$) and the tensor density
  $\alpha^{(N\Lambda)}_{T}\rho^\Lambda_T(\br)$ (c)
with respect to the maximum number of major shell ($N_f$) in the 3DHO basis for
the triaxial RMF calculation for $^{17}_\Lambda$O. These quantities are
evaluated at $x=y=0.36$ fm and plotted as a function of $z$.
For a comparison, the results of spherical RMF calculations are also shown
by the lines.}
\label{fig:rhot}
\end{figure}

 \section{Results of mean-field studies}
 \label{Sec.III}
  \subsection{Illustrative calculations}

  We first carry out an illustrative calculation to test
  our triaxial RMF code for some spherical hypernuclei, comparing with the
  results by the spherical RMF code in coordinate space with the box size
  of $R=15$ fm~\cite{Tanimura2012_PRC85-014306}. If not particularly indicated, the same parameters
  are used in both calculations, including the masses of nucleons
  and $\Lambda$ which are taken as $m_N$ = 939 MeV$/c^2$ and $m_\Lambda$
  = 1115.6 MeV$/c^2$, respectively and the  PC-F1 force
  \cite{Burvenich2002_PRC65-044308} for the $NN$ interaction and the
  PCY-S1 \cite{Tanimura2012_PRC85-014306} for the $N\Lambda$ interaction.
Taking $^{17}_{\Lambda s}$O as an example, where the $\Lambda$ particle is put in the
$1s_{1/2}$ orbital, we plot in Fig.~\ref{fig:rhot} the mean-field potentials
$V^B_0+S^B$ and $V^B_0-S^B$ for neutron, proton and $\Lambda$,
together with the tensor density $\alpha^{(N\Lambda)}_{T}\rho^\Lambda_T$
from the triaxial RMF calculation. We compare the results with three
different maximum major-shell numbers in the 3DHO basis, $N_f$=8, 10, and
12. The results are compared also with those by the spherical RMF calculations.
It is shown that the mean-field potentials are well converged
at $N_f=10$. In contrast, the tensor density (\ref{rhotensor})
originated from the tensor coupling term ${\cal L}^{N\Lambda}_{\rm ten}$
has a slightly slower convergence behavior, especially at small
values of $z$. It is shown that the tensor density obtained with the
maximum shell number $N_f=12$ gives a good agreement with
the spherical RMF result.

\begin{table*}[t]
\tabcolsep=6pt
\caption{The total energy $E_{\textrm{tot}}$, the
  kinetic energy $E_{\textrm{kin}}(=T_N+T_\Lambda)$, the root-mean-square (rms)
  radii of neutrons $r_n$, protons $r_p$ and hyperon $r_\Lambda$,
  and the energy of the lowest three single-particle states for neutron,
  proton and hyperon obtained with the triaxial RMF (Tri.RMF) calculation
  for $^{17}_{\Lambda s}$O, $^{31}_{\Lambda s}$Si, $^{33}_{\Lambda s}$S,
  and $^{41}_{\Lambda s}$Ca, in comparison with those with the spherical
  RMF (Sph.RMF) calculations. All the energies are in MeV and the
  radii are in fm.}
\label{table1}
\vspace{0.2cm}
\centering
\begin{tabular}{ccccccccccccccccccc}
\hline \hline
&     &    \multicolumn{2}{c}{$^{17}_{\Lambda s}$O}   &\multicolumn{2}{c}{$^{31}_{\Lambda s}$Si}   & \multicolumn{2}{c}{$^{33}_{\Lambda s}$S}  & \multicolumn{2}{c}{$^{41}_{\Lambda s} $Ca } \\
 \cline{3-4\  \ } \cline{5-6\  \ } \cline{7-8\  \ }\cline{9-10\  \ }
& &  Sph.RMF & Tri.RMF &  Sph.RMF & Tri.RMF &  Sph.RMF & Tri.RMF &  Sph.RMF & Tri.RMF \\
\hline
& $E_{\textrm{tot}}$      & $-$140.317  & $-$140.309 & $-$269.491 &  $-$269.476  & $-$285.434 & $-$285.320 & $-$363.459 & $-$363.174 \\
& $E_{\textrm{kin}}$      & 210.036   & 210.031  &  435.385 &  435.436 &  439.216 & 439.164  & 518.686  & 518.239 \\
& $E_{\textrm{cm}}$       & 9.752     & 9.750    & 9.282    &  9.281 & 8.915    & 8.902    & 8.167    & 8.140 \\
& $r_n$                   & 2.613     & 2.613    & 3.084    &  3.083  & 3.088    & 3.088    & 3.340    & 3.341 \\
& $r_p$                   & 2.638     & 2.638    & 2.984    &  2.984  & 3.129    & 3.129    & 3.385    & 3.386 \\
& $r_\Lambda$             & 2.458     & 2.458    & 2.516    &  2.515 & 2.571    & 2.570    & 2.820    & 2.823 \\
\hline
$\multirow{3}{*}{\centering neutron}$ & 1s$_{1/2}$ & $-$41.629 & $-$41.628 & $-$54.234 &  $-$54.260 & $-$57.544 & $-$57.528 & $-$53.827 & $-$53.817  \\
                                      & 1p$_{3/2}$ & $-$21.937 & $-$21.937 & $-$34.531 &  $-$34.551 & $-$36.477 & $-$36.480 & $-$37.859 & $-$37.854  \\
                                      & 1p$_{1/2}$ & $-$15.285 & $-$15.288 & $-$27.694 &  $-$27.728 & $-$28.480 & $-$28.496 & $-$33.354 & $-$33.362  \\ \\

$\multirow{3}{*}{\centering proton}$  & 1s$_{1/2}$ & $-$37.517 & $-$37.514 & $-$51.510  & $-$51.521  & $-$50.462 & $-$50.445 & $-$45.721 & $-$45.712 \\
                                      & 1p$_{3/2}$ & $-$18.107 & $-$18.105 & $-$30.474  &  $-$30.490 & $-$29.744 & $-$29.746 & $-$30.102 & $-$30.096 \\
                                      & 1p$_{1/2}$ & $-$11.531 & $-$11.533 & $-$23.848 & $-$23.877 & $-$21.827 & $-$21.841 & $-$25.601 & $-$25.608 \\\\

$\multirow{3}{*}{\centering hyperon}$ & 1s$_{1/2}$ & $-$12.569 & $-$12.570 & $-$18.908 & $-$18.946 & $-$18.458 & $-$18.483 & $-$18.305 & $-$18.278 \\
                                      & 1p$_{3/2}$ & $-$2.336  & $-$2.297  &  $-$8.485  &  $-$8.510 & $-$8.405  & $-$8.416  & $-$10.112 & $-$10.108 \\
                                      & 1p$_{1/2}$ & $-$1.995  & $-$1.947  &  $-$8.391   &  $-$8.406 & $-$8.189  & $-$8.207  & $-$10.255 & $-$10.263 \\
\hline \hline
\end{tabular}
\end{table*}

Table~\ref{table1} shows the detailed structural properties of
hypernuclei,  $^{17}_{\Lambda s}$O, $^{31}_{\Lambda s}$Si, $^{33}_{\Lambda s}$S,
and $^{41}_{\Lambda s}$Ca from both the spherical and triaxial RMF calculations.
In the triaxial RMF calculation, fourteen major HO shells ($N_f$=14) are
adopted to expand the Dirac spinors, with which the quadrupole deformation $\beta$
of hypernuclei is convergent to zero. It is shown that both approaches
give the results very close to each other.  The remaining small
differences in the binding energies can be reduced further by
increasing the  maximum major-shell number $N_f$ of the HO basis and
constraining the high-order hexadecapole deformation to be zero in
the triaxial RMF calculation.

We note that the results presented in Tab.~\ref{table1} are obtained without breaking the symmetry of time-reversal invariance in the mean-field calculations for the $\Lambda$ hypernuclei. It has been found in recent studies~\cite{Lijian2009,Xu2014} that the effect of time-odd fields from the breaking of time-reversal invariance by one unpaired nucleon on the binding energies
is in between 0.1 and 0.2 MeV for the ground state of
the sd-shell odd-mass nuclei.
For the single-$\Lambda$ hypernuclei with an even-even nuclear core and the $\Lambda$ in a low orbital-angular-momentum state, the effect caused by the unpaired $\Lambda$ hyperon is even smaller and it leads to an energy splitting of 0.1 MeV for the time-reversal partner states of $s_{1/2}$ orbital in $^{17}_\Lambda$O~\cite{Sang2013.PhysRevC.88.064304}. Since the time-reversal invariance is imposed in both the spherical and triaxial deformed RMF calculations, this effect will not contribute to the difference in the two results.

 \subsection{Hyperon binding energy and deformation effect}

\begin{figure}[h!]
\centering
\includegraphics[width=9cm]{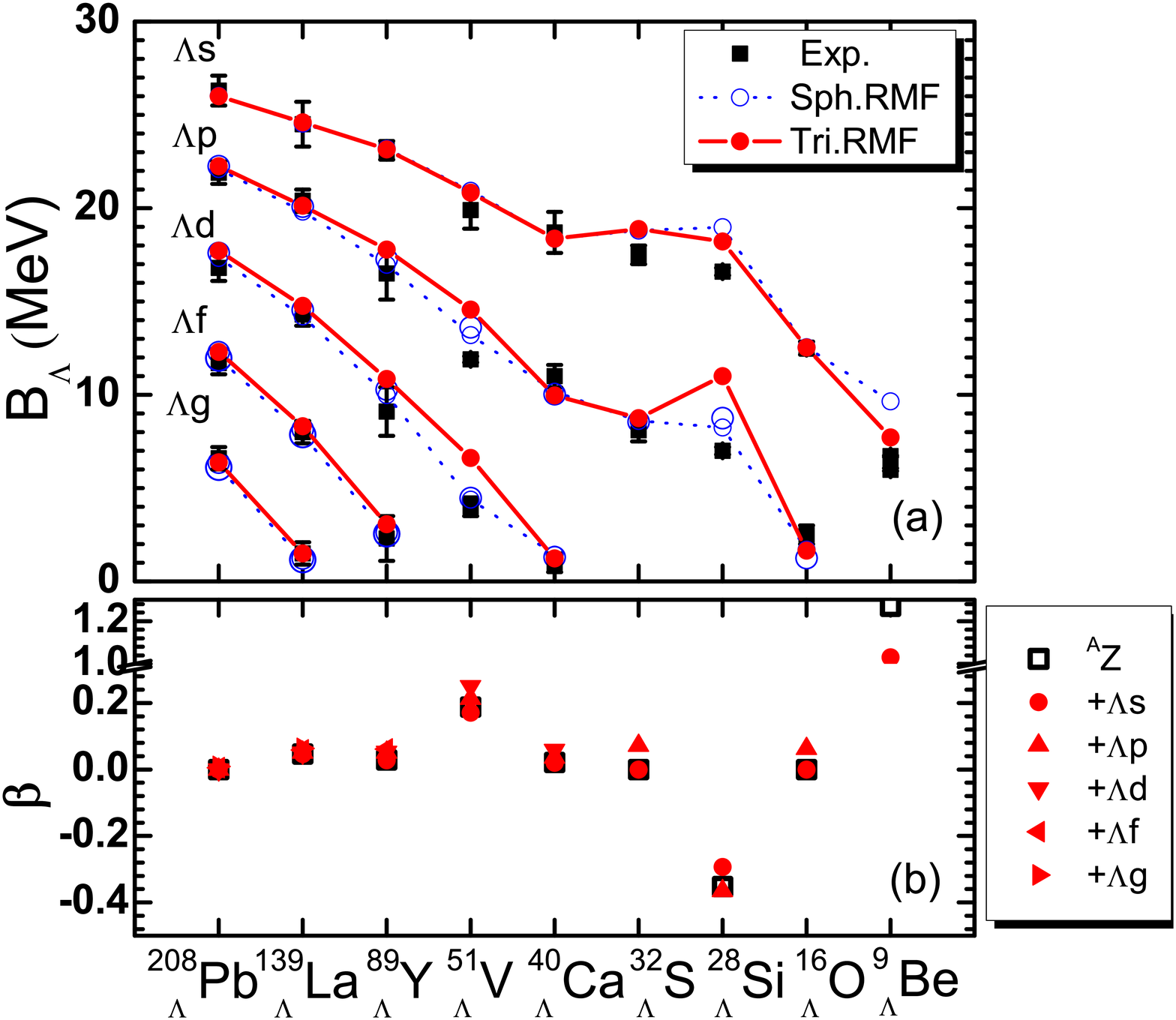}
\caption{(Color online) (a) The $\Lambda$  binding energies in
  single-$\Lambda$ hypernuclei with $\Lambda$ in different single-particle
  states obtained with the spherical and the triaxial RMF calculations,
  and their comparison with the available data from Refs.
\cite{Hotchi2001PhysRevC.64.044302,Hashimoto2006_PPNP-564,
Usmani1999_PRC60-055215,Motoba1998_NPA-135c}.
The parameter set PC-F1 and PCY-S1 are adopted for the $NN$
and the $N\Lambda$ interactions, respectively.
(b) The quadrupole deformation $\beta$  for each $\Lambda$ hypernuclei
obtained with the triaxial RMF calculation. See text for more details.}
\label{fig:B_L}
\end{figure}

\begin{figure}[]
\centering
\includegraphics[width=8cm]{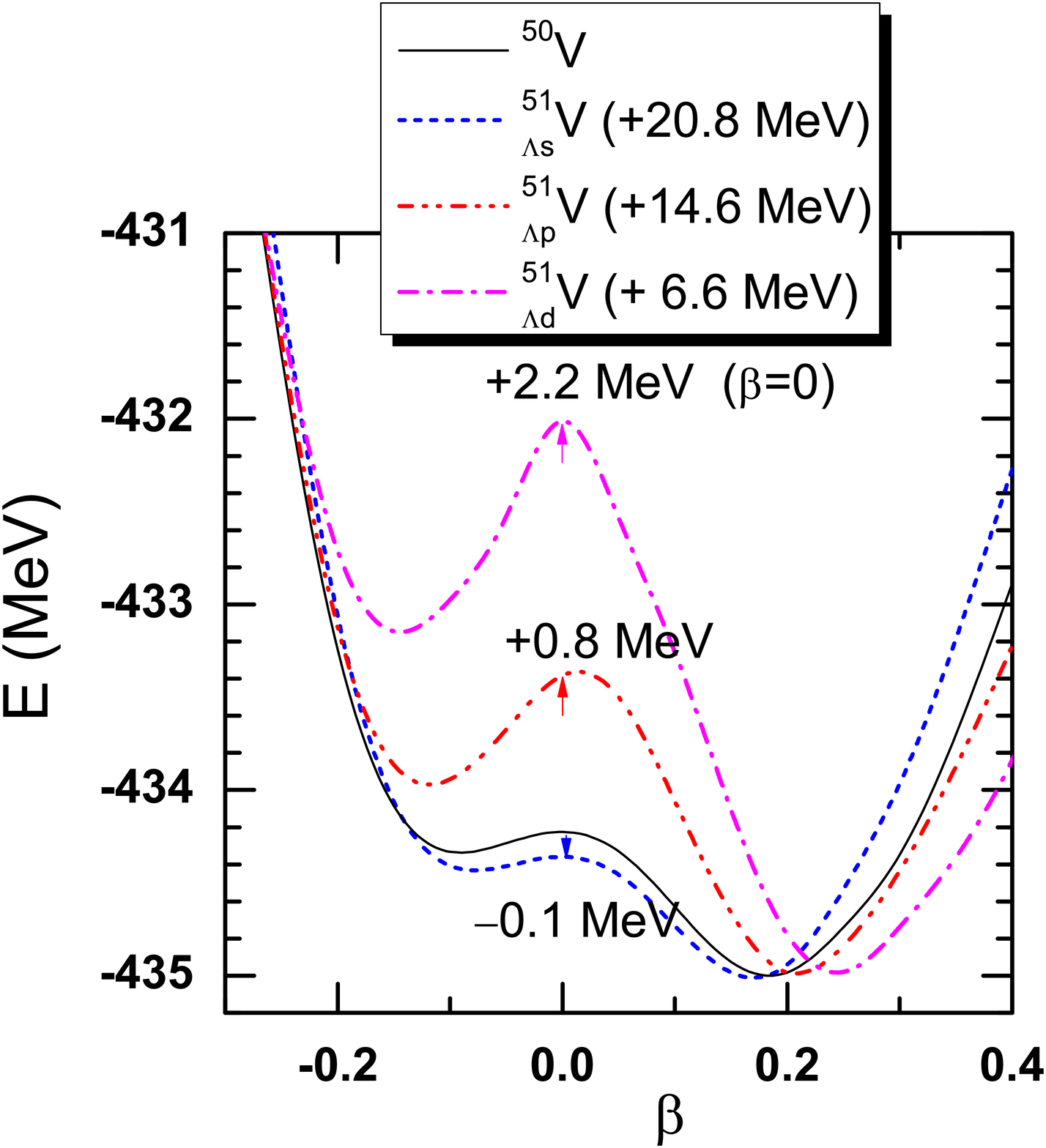}
\caption{(Color online) The total energy of $^{50}$V, $^{51}_{\Lambda s}$V,
  $^{51}_{\Lambda p}$V and $^{51}_{\Lambda d}$V as a function of deformation
  parameter $\beta$ with the triaxially deformed RMF calculations.
  The energy of hypernuclei is shifted by normalizing the minimum
  energy to that of $^{50}$V. The energy difference between the hypernuclei
  and $^{50}$V  at $\beta=0$ is indicated with the numbers.}
\label{fig:V50}
\end{figure}

We first discuss the effect of deformation on the binding energy of hypernuclei.
Figure~\ref{fig:B_L}(a) shows the $\Lambda$ binding energies $B_\Lambda$
in single-$\Lambda$ hypernuclei obtained with the spherical and triaxial RMF
calculations, in comparison with the experimental data.
Here, $B_\Lambda$ is defined as the energy difference between the ground
state (g.s.) of hypernucleus and that of the core nucleus, that is,
$B_\Lambda\equiv E_{\rm tot}(^AZ, g.s.)-E_{\rm tot}(^{A+1}_{~~~\Lambda} Z, g.s.)$.
In the triaxial RMF calculation for the $\Lambda$ hypernuclei,
the $\Lambda$ hyperon is always put in the lowest state among those
which are connected to the $s, p, $\ldots$, g$ state in the spherical limit.
The corresponding $\Lambda$ is therefore denoted
as $\Lambda_s$, $\Lambda_p$, $\ldots$, $\Lambda_g$ for convenience.
In the spherical RMF calculations, two values of $B_\Lambda$ for
the $\ell_\Lambda \neq 0$ cases are plotted, corresponding to the
$\Lambda$ hyperon in the spin-orbit partner states.
Due to the introduction of a strong $\Lambda$ tensor coupling,
the energy splitting of the spin-orbit partner states by the
spherical RMF calculation is less than 0.5 MeV and mostly with
an opposite sign to the ordinary nuclei~\cite{Tanimura2012_PRC85-014306}.
The triaxial RMF approach yields similar results as the spherical
RMF approach for the $\Lambda$ binding energies of most hypernuclei,
except for $^{9}_\Lambda$Be, $^{28}_{~\Lambda}$Si and $^{51}_{~\Lambda}$V which
are deformed in their ground states (See Fig.~\ref{fig:B_L}(b) for the
value of deformation parameter).
For these nuclei,
a nonzero deformation $\beta$ decreases the binding energy
of $\Lambda_s$ and improves the agreement with the data, while it
signinificantly overestimates the binding energy of $\Lambda_p$
and $\Lambda_d$.

We take hypernucleus $^{51}_{~\Lambda}$V as an example to illustrate the
deformation effect on $B_\Lambda$. Figure~\ref{fig:V50} shows that
the energy minimum of hypernucleus $^{51}_{\Lambda s}$V is shifted
slightly towards spherical shape, while that of $^{51}_{\Lambda p}$V
and $^{51}_{\Lambda d}$V is pushed to a larger deformed shape.
Moreover, it is shown that the deformation of hypernuclei increases
from $^{51}_{\Lambda s}$V to $^{51}_{\Lambda p}$V, and then to $^{51}_{\Lambda d}$V.
The difference in the $B_\Lambda$ values of $^{51}_\Lambda$V by
the spherical and triaxial RMF calculations is also shown
clearly in Fig.~\ref{fig:V50}, where the energy of hypernucleus
$^{51}_{~\Lambda}$V decreases or increases by 0.1 MeV, 0.8 MeV and 2.2 MeV
for the $\Lambda_s$, $\Lambda_p$ and $\Lambda_d$, respectively.
The microscopic mechanism responsible for these  phenomena
can be traced to the Nilsson diagram of hyperon single-particle
energies, which will be discussed in details in the next subsection.

 \subsection{Shape polarization effect of $\Lambda$ hyperon in $sd$-shell nuclei in $(\beta, \gamma)$ plane}

\begin{figure}[]
\centering
\includegraphics[width=9cm]{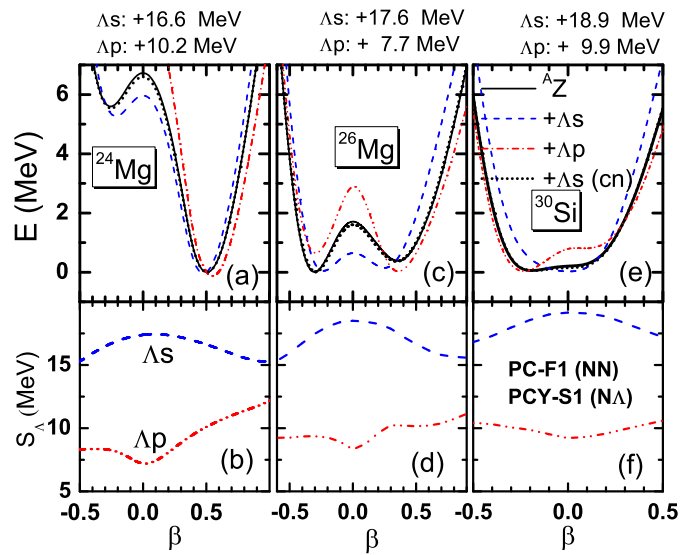}
\caption{(Color online) The potential energy surfaces
obtained with the triaxially deformed RMF calculations for
(a)  $^{24}$Mg, $^{25}_{\Lambda s}$Mg, $^{25}_{\Lambda p}$Mg and the
  core nucleus inside $^{25}_{\Lambda s}$Mg; (c) $^{26}$Mg, $^{27}_{\Lambda s}$Mg,
  $^{27}_{\Lambda p}$Mg and the core nucleus inside $^{27}_{\Lambda s}$Mg;
  (e) $^{30}$Si, $^{31}_{\Lambda s}$Si, $^{31}_{\Lambda p}$Si and the
  core nucleus inside $^{30}_{\Lambda s}$Si  as a function of deformation
  parameter $\beta$. All the energies are normalized to the global minimum. The deformation-dependent
  $\Lambda$ separation energy defined by Eq. (\ref{s-lambda}) as a function of deformation parameter $\beta$ for
(b) $^{25}_{\Lambda s}$Mg and $^{25}_{\Lambda p}$Mg; (d) $^{27}_{\Lambda s}$Mg and $^{27}_{\Lambda p}$Mg; and (f) $^{31}_{\Lambda s}$Si and $^{31}_{\Lambda p}$Si.  The binding energies of $\Lambda_s$ and $\Lambda_p$ for each case are given on the top of the figures.}
\label{fig:Lambdadeform}
\end{figure}

In order to discuss more in details the shape polarization effect of $\Lambda$ hyperon,
we take as examples three  $sd$-shell hypernuclei,
the prolate deformed $^{24}$Mg and the oblate deformed $^{26}$Mg
and $^{30}$Si, the latter two having transitional characters.
Figure~\ref{fig:Lambdadeform} (a) shows the PESs for
the $^{25}_{\Lambda s}$Mg, $^{25}_{\Lambda p}$Mg, $^{27}_{\Lambda s}$Mg,
$^{27}_{\Lambda p}$Mg, $^{31}_{\Lambda s}$Si, and $^{31}_{\Lambda p}$Si obtained with
the triaxial RMF method as a function of deformation parameter $\beta$.
The energy surfaces for the corresponding core nuclei are also shown.
As in the case of $^{51}_{~\Lambda}$V shown in Fig.~\ref{fig:V50},
the energy minimum of $^{25}_{\Lambda s}$Mg and $^{25}_{\Lambda p}$Mg
is shifted to a slightly smaller and larger deformed region, respectively,
compared with that of the core nucleus.
For $^{26}$Mg, the $\Lambda_s$ significantly lowers down
the barrier between the oblate and prolate minima.
Of particular interest is that the $\Lambda_p$ inverts the
energy order of the oblate and prolate minima in $^{26}$Mg.
A significant change of the deformation parameter of the
mean-field ground state by $\Lambda$ hyperon is shown in $^{30}$Si.
That is, the $\Lambda_s$ brings the oblate deformed $^{30}$Si to
spherical $^{31}_{\Lambda s}$Si.
A similar conclusion has been obtained also in Refs.
~\cite{WH08,Lu2011_PRC84-014328}. In contrast,
the $\Lambda_p$ drives $^{30}$Si to be more oblate deformed.
It is worthwhile to mention that although the PESs of hypernuclei
could be significantly different from those of core nuclei,
the differences in the PESs for the core nuclei inside the
hypernuclei (the dotted lines) and for the core nuclei
without the hyperon impurity (the solid lines) are negligibly small.

To illustrate the shape-driving effect of $\Lambda$ hyperon in different orbitals, we introduce a deformation-dependent
$\Lambda$ separation energy $S_\Lambda(\beta)$ as
\begin{eqnarray}
\label{S_L}
S_\Lambda (\beta)  &\equiv& E_{\rm tot}(^AZ, \beta)-E_{\rm tot}(^{A+1}_{~~~\Lambda} Z, \beta),
\label{s-lambda}
\end{eqnarray}
where $E_{\rm tot}(^{A+1}_{~~~\Lambda} Z, \beta)$ and $E_{\rm tot}(^AZ, \beta)$
are the total energies of the hypernucleus and the core nucleus at
deformation $\beta$, respectively.
Here, we take the same deformation value $\beta$ for the hypernucleus
and the core nucleus to define the quantity $S_\Lambda(\beta)$.
Notice that even though $S_\Lambda(\beta)$ is
different from the standard definition of hyperon separation energy, it
provides a convenient way to understand the shape polarization
effect of $\Lambda$. The $S_\Lambda (\beta)$ in  $^{25}_{\Lambda s}$Mg,
$^{25}_{\Lambda p}$Mg, $^{27}_{\Lambda s}$Mg, $^{27}_{\Lambda p}$Mg, $^{31}_{\Lambda s}$Si,
and $^{31}_{\Lambda p}$Si  is shown
in Figs.~\ref{fig:Lambdadeform} (b), (d) and (f)
as a function of deformation $\beta$.
It is seen clearly that the $S_\Lambda$ decreases (or increases)
with $\beta$ for $\Lambda_s$ (or $\Lambda_p$), which provides
a mechanism to change the structure of the PESs in such a way
that the nuclear shape becomes less or more deformed.

\begin{figure}[]
\centering
\includegraphics[width=7cm]{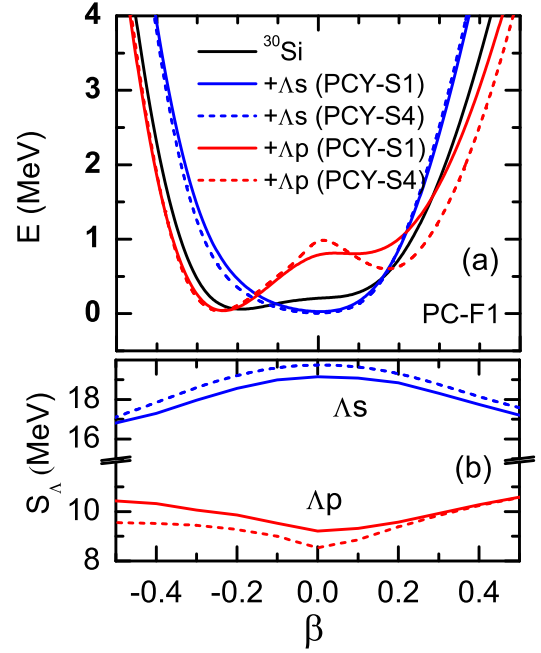}
\caption{(Color online) (a) The potential energy surfaces for $^{30}$Si, $^{31}_{\Lambda s}$Si, $^{31}_{\Lambda p}$Si as a function of deformation  parameter $\beta$  obtained with the deformed RMF calculations using the same PC-F1 force for the $NN$ interaction but different $N\Lambda$ interaction (PCY-S1 and PCY-S4~\cite{Tanimura2012_PRC85-014306}, respectively). The energies are normalized to the global minima. (b) The $\Lambda$ separation energy $S_\Lambda (\beta)$ as a function of deformation parameter $\beta$.}
\label{fig:30Si-comp}
\end{figure}

Figure~\ref{fig:30Si-comp}(a) shows a
comparison of the potential energy surfaces for $^{30}$Si, $^{31}_{\Lambda s}$Si, and $^{31}_{\Lambda p}$Si from the deformed RMF calculation with different $N\Lambda$ interactions, namely the PCY-S1 and PCY-S4~\cite{Tanimura2012_PRC85-014306}. It is seen that the potential energy surface
of $^{31}_{\Lambda s}$Si is similar to each other for both of
the two $N\Lambda$ interactions. However, that of $^{31}_{\Lambda p}$Si are evidently different in some deformation region. This difference is shown more clearly in the comparison of $\Lambda$ separation energy $S_\Lambda (\beta)$ in Fig.~\ref{fig:30Si-comp}(b). For $\Lambda_s$, the two $N\Lambda$ interactions give similar slops of $S_\Lambda(\beta)$ as a function of deformation $\beta$. On the other hand,
for $\Lambda_p$, the slop is apparently different on the prolate side, resulting in the different behavior of the potential energy surface of $^{31}_{\Lambda p}$Si. It implies that the impurity effect of $\Lambda_p$ is somewhat sensitive to the $N\Lambda$ interaction, in contrast to that of $\Lambda_s$.

\begin{figure}[tb]
\centering
\includegraphics[width=8cm]{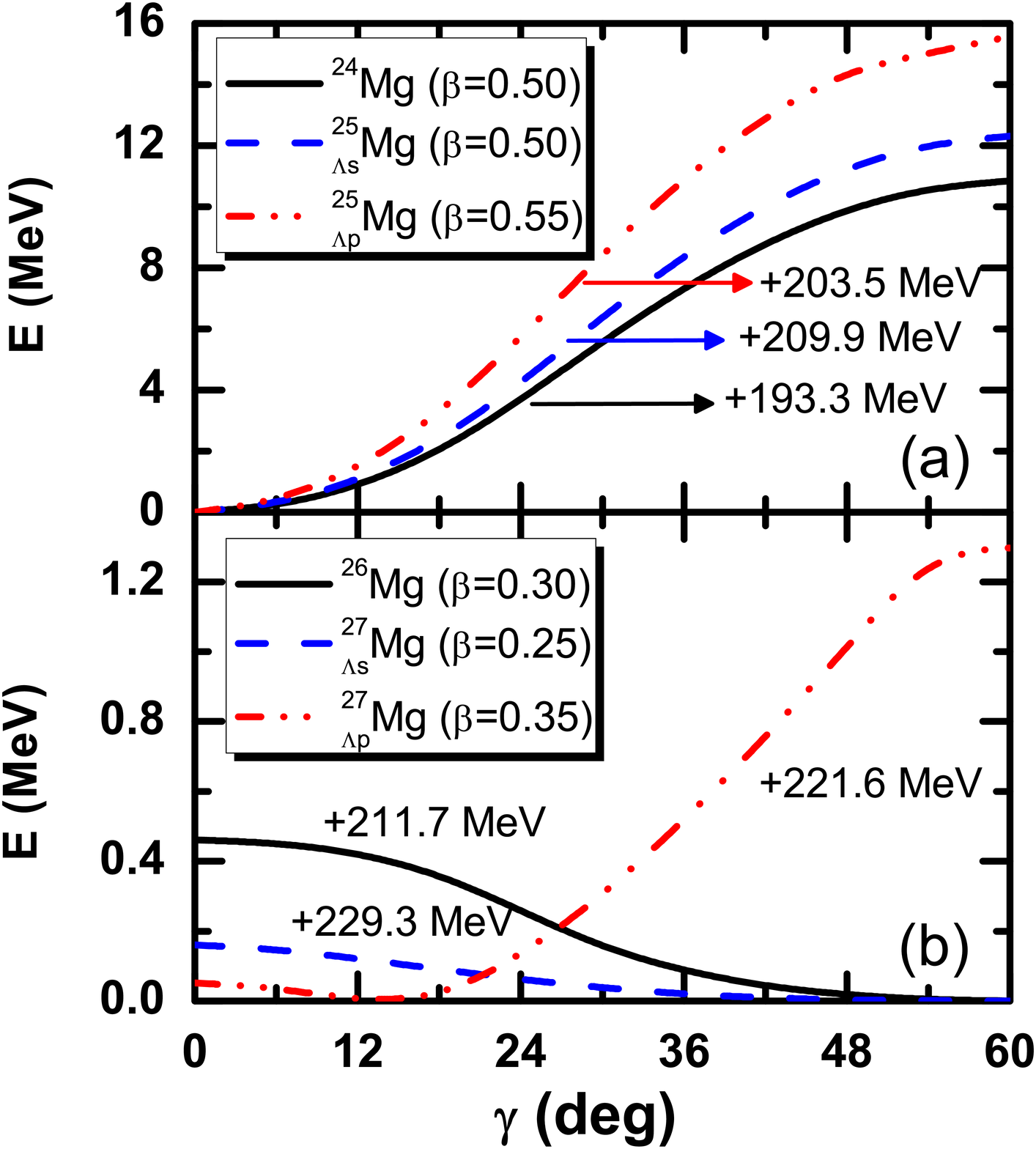}
\caption{(Color online) (a) The total energy of $^{24}$Mg,
  $^{25}_{\Lambda s}$Mg and $^{25}_{\Lambda p}$Mg
  as a function of deformation parameter $\gamma$,
  where the deformation $\beta$ is fixed at the value of the
  global minimum of the energy surface.
  The energies are normalized to the energy of the global minimum.
  (b) Same as (a), but for $^{26}$Mg, $^{27}_{\Lambda s}$Mg and
  $^{27}_{\Lambda p}$Mg.}
\label{fig:gamma}
\end{figure}

The influence of $\Lambda$ hyperon on nuclear triaxiality is
illustrated in Fig.~\ref{fig:gamma}, which shows the PESs for
the $^{25}_{\Lambda s}$Mg, $^{25}_{\Lambda p}$Mg,
$^{27}_{\Lambda s}$Mg, and $^{27}_{\Lambda p}$Mg, and their core nuclei
as a function of $\gamma$ deformation. For $^{24}$Mg with a pronounced
prolate energy minimum, the stiffness of the PES along $\gamma$
deformation increases by adding $\Lambda_s$ or $\Lambda_p$.
This is an opposite tendency from that predicted by the Skyrme-Hartree-Fock
(SHF) calculation \cite{WHK11}.
For $^{26}$Mg, with a shallow oblate energy minimum,
the inclusion of $\Lambda_s$ softens the PES along $\gamma$ deformation,
in agreement with the previous SHF calculation. However, at the same time,
the energy minimum is shifted to a prolate-like shape with $\gamma=12^\circ$
by adding $\Lambda_p$, which is a consequence of competition
between the $\gamma$ deformation effect and the $\Lambda_p$ impurity effect.

 \begin{figure}
\centering
\includegraphics[width=8cm]{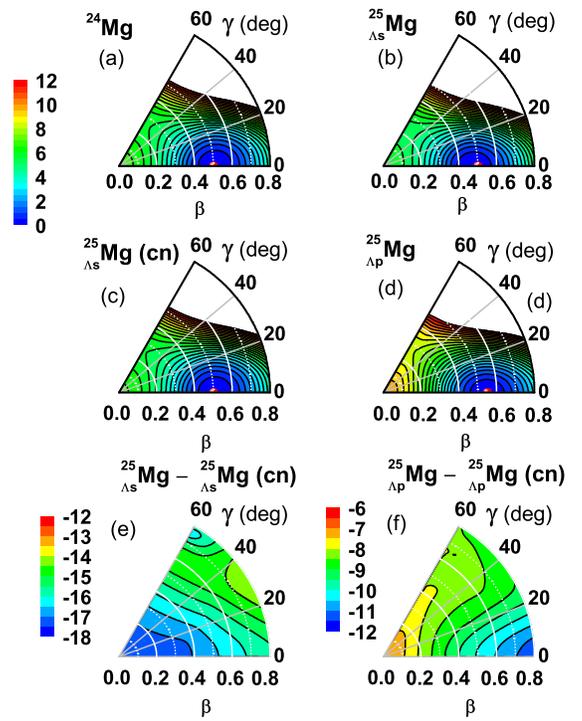}\vspace{-1.5cm}
\caption{(Color online) The potential energy surfaces of
  $^{24}$Mg (a), $^{25}_{\Lambda s}$Mg (b), $^{25}_{\Lambda s}$Mg (cn) (c),
  $^{25}_{\Lambda p}$Mg (d) in the $(\beta, \gamma)$ plane.
  The energies are normalized to the global minimum.  The energy
  difference between $^{25}_{\Lambda s}$Mg and its core nucleus (e), and
  that between $^{25}_{\Lambda p}$Mg and its core nucleus (f) are also
  plotted. Two neighboring contour lines are separated by 0.5 MeV. }
\label{fig:PES24Mg}
\end{figure}
\begin{figure}
\centering
\includegraphics[width=8cm]{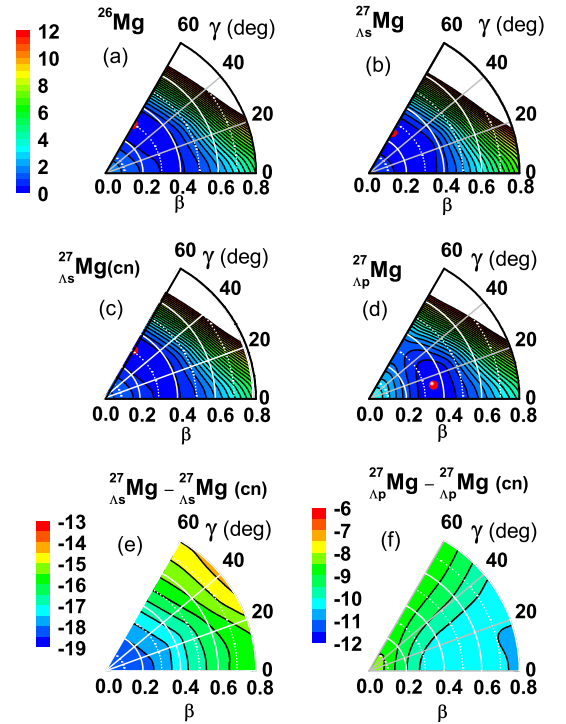}\vspace{-1.5cm}
\caption{(Color online) Same as Fig. \ref{fig:PES24Mg}, but for
  $^{26}$Mg, $^{27}_{\Lambda s}$Mg, and $^{27}_{\Lambda p}$Mg.}
\label{fig:PES26Mg}
\end{figure}
 \begin{figure}
\centering
\includegraphics[width=8cm]{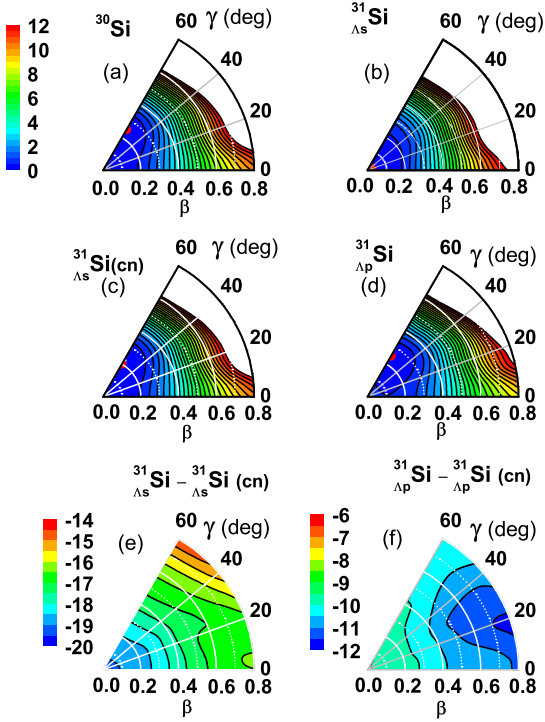}\vspace{-1.5cm}
\caption{(Color online) Same as Fig. \ref{fig:PES24Mg}, but for
  $^{30}$Si, $^{31}_{\Lambda s}$Si and $^{31}_{\Lambda p}$Si.}
\label{fig:PES30Si}
\end{figure}

 The impurity effects of $\Lambda_s$ and $\Lambda_p$ on nuclear
 quadrupole deformation $\beta$ and $\gamma$ discussed in Figs.~\ref{fig:Lambdadeform} and \ref{fig:gamma}
can be also studied in the PES in the whole $(\beta, \gamma)$ plane.
Figs.~\ref{fig:PES24Mg}, \ref{fig:PES26Mg} and \ref{fig:PES30Si} show
the PESs for $^{24,26}$Mg and $^{30}$Si, together with the hypernuclei
$^{25,27}_{~~~~\Lambda}$Mg and $^{31}_{~\Lambda}$Si with a $\Lambda_s$ or $\Lambda_p$.
The contribution of $\Lambda_s$ and $\Lambda_p$ to the total energy of
hypernuclei in $(\beta, \gamma)$ plane is also plotted. One can again
see that the $\Lambda_s$ stabilizes the spherical shape,
while the $\Lambda_p$ stabilizes deformed shape.

\begin{figure}
\centering
\includegraphics[width=8.5cm]{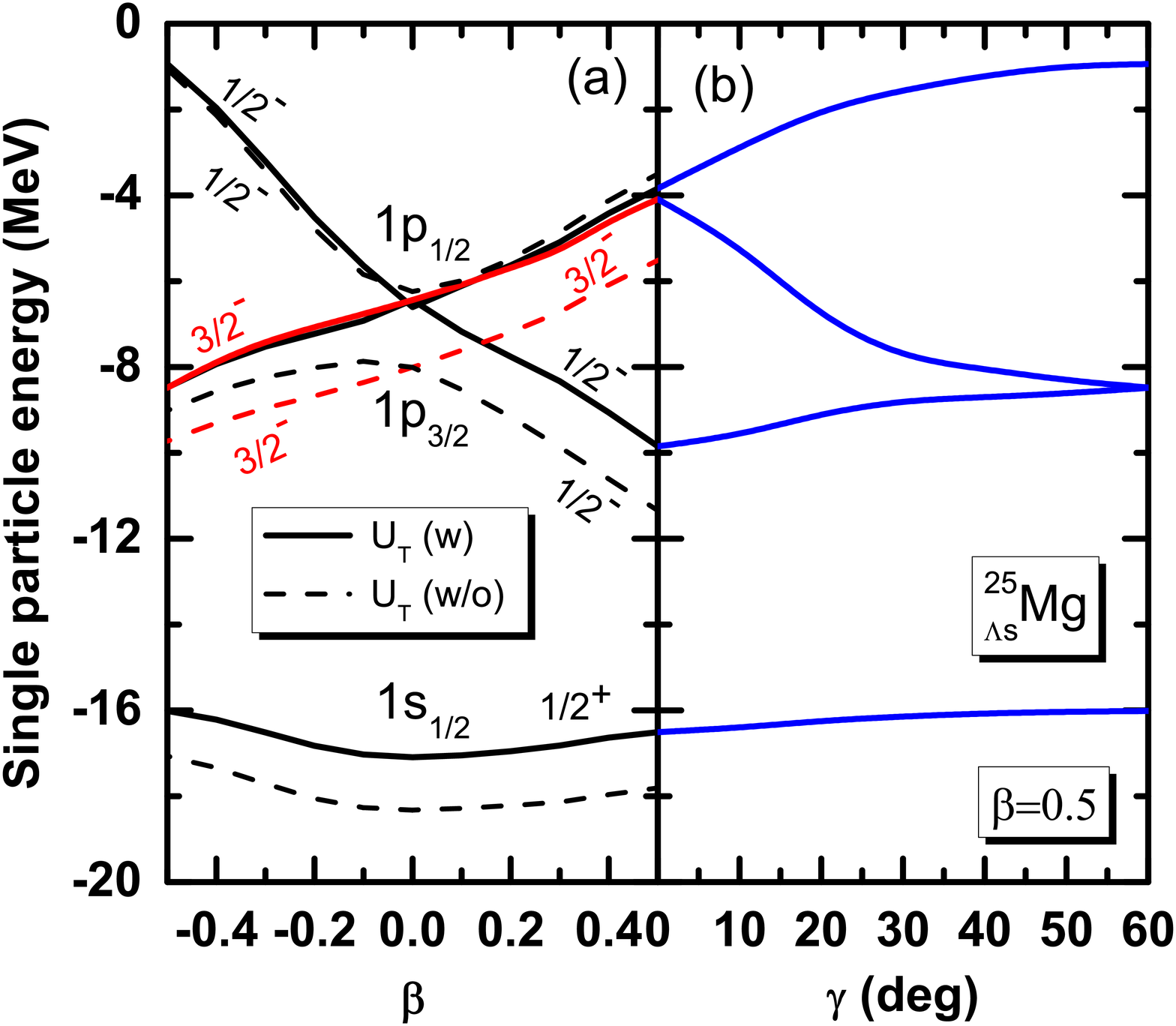}
\caption{(Color online) The single-particle energies of
  $\Lambda$ hyperon in $^{25}_{\Lambda s}$Mg as a function of deformation
  parameters $\beta$ and $\gamma$ from the calculation with the PC-F1 and PCY-S1 for the $NN$ and $N\Lambda$ interactions, respectively. In the panel (a), the energy levels from the calculation without the tensor potential $U_T$ (\ref{U_T}) are also plotted with the dashed lines.}
\label{fig:spe}
\end{figure}

To understand the impurity effects of $\Lambda_s$ and $\Lambda_p$
in a qualitative way, we plot in Fig.~\ref{fig:spe}
the Nilsson diagram of the single-particle energy for $\Lambda$ hyperon
in $^{25}_{\Lambda s}$Mg as a function of deformation parameters
$\beta$ and $\gamma$. The results of the calculation without the tensor potential $U_T$
in the Dirac equation for $\Lambda$ hyperon are also plotted
for comparison. It is shown that the tensor potential $U_T$ pushes
up the energy of 1$s_{1/2}$ and 1$p_{3/2}$ orbitals and reduces
significantly the spin-orbit splitting between the
partner states $\Lambda p_{3/2}$ and $\Lambda p_{1/2}$ at the spherical shape.
The resultant splitting energy is $\Delta E_{\rm so}=\epsilon_{j=l-1/2}-\epsilon_{j=l+1/2}=-0.16$ MeV.
The inversion of the energy order of the spin-orbit partner states
for $\Lambda$ hyperon is a particular character of the parameter set
PCY-S1 for $N\Lambda$ interaction with a very strong tensor
coupling~\cite{Tanimura2012_PRC85-014306}.

With the increase of deformation $\beta$, the $\Lambda_s$
becomes slightly less bound, while the $\Lambda_p$ becomes deeper bound.
It explains both the behaviors of $\Lambda$ separation energy as a
function of $\beta$ and the different shape-driving effect
of $\Lambda_s$ and $\Lambda_p$ shown in Fig.~\ref{fig:Lambdadeform}.
In other words, the $\Lambda_s$ drives the hypernucleus
towards spherical shape, while the $\Lambda_p$ drives the hypernucleus
towards large deformation, as has been pointed out in Ref.~\cite{IKDO11}
based on the AMD calculations.
For the axial asymmetric shapes,
the energies of the three $p$-hyperon orbitals in $^{25}_{\Lambda s}$Mg
are apparently different from each other. The rotational bands with
such configurations have been discussed recently
based on the AMD model~\cite{Isaka13}.
The energy of $\Lambda_p$ increases with $\gamma$,
explaining the phenomenon that the $\Lambda_p$ drives hypernucleus
towards prolate-like shape with slightly larger
deformation $\beta$ (see Fig.~\ref{fig:gamma} (b)).

 \section{Projected potential energy surfaces for $^{25}_\Lambda$Mg with microscopic particle-rotor model}
 \label{Sec.IV}

 The deformed mean-field states considered in the previous section
 break rotational symmetry and thus several angular momentum components
 are admixed in the wave functions.
 To compare with experimental data, one has to make a transformation
 from the intrinsic frame to the laboratory frame, which can be realized
 by introducing the technique of AMP. It can be implemented  based on the mean-field wave function for the
 whole $\Lambda$ hypernuclei composed of an even-even nuclear core and
 one unpaired $\Lambda$ particle. Rather than implementing the AMP
 for odd-mass nuclear system, however, we instead apply the microscopic
 particle-rotor model (PRM)~\cite{MHYM14} to calculate the
 PESs for the hypernuclei.

 The microscopic PRM was developed recently by the present authors
 using the transition
 densities from the multi-reference DFT calculation~\cite{Yao13PLB,Yao14TD}.
 In this model, the wave function of $\Lambda$ hypernucleus is constructed as
 \begin{equation}
 \label{wavefunction}
 \displaystyle \Psi_{IM}(\br_{\Lambda},\{\br_N\})
 =\sum_{j\ell J}  {\mathscr R}_{j\ell J}(r_{\Lambda}) {\mathscr F}^{IM}_{j\ell J}(\hat{\br}_{\Lambda}, \{\br_N\}),
\end{equation}
where ${\mathscr F}^{IM}_{j\ell J}$ is given by
\begin{equation}
 {\mathscr F}^{IM}_{j\ell J}(\hat{\br}_{\Lambda}, \{\br_N\})
= [{\mathscr Y}_{j\ell}(\hat{\br}_{\Lambda})\otimes
\Phi_{J}(\{\br_N\})]^{(IM)}
\end{equation}
with $\br_{\Lambda}$ and $\br_N$ being the coordinates of the $\Lambda$ hyperon and the
nucleons, respectively.  Here,
$I$ is the total angular momentum and $M$ is its projection onto the
$z$-axis for the whole $\Lambda$ hypernucleus.
${\mathscr R}_{j\ell J}(r_{\Lambda})$
and ${\mathscr Y}_{j\ell}(\hat{\br}_{\Lambda})$ are the four-component radial wave function and
the spin-angular wave function for the $\Lambda$ hyperon, respectively.
In this paper,
the wave function $\Phi_{J}(\{\br_N\})$ for the nuclear core is chosen
as the projected mean-field wave function with different intrinsic
deformation $\beta$, that is,
 \begin{equation}
 \vert \Phi_{JM_J} (\beta)\rangle
 =  \hat P^{J}_{M_JK} \hat P^N\hat P^Z\vert \varphi(\beta)\rangle,
\label{GCM}
 \end{equation}
 where $\hat P^{J}_{M_JK}$ is the projection operator onto a good number
 of angular momentum, while
$\hat P^N$ and $\hat P^Z$ are those for
 neutron and proton numbers, respectively.
 The total energy of a $\Lambda$ hypernucleus with spin-parity
 $I^\pi$ corresponding to deformation $\beta$ of the core nucleus is
 defined as the energy $E_{I}$ of the lowest solution
 of the equation $\hat H|\Psi_{IM}\rangle=E_{I}|\Psi_{IM}\rangle$, from which
 one can derive the coupled-channels equations for
 ${\mathscr R}_{j\ell J}(r_{\Lambda})$.
   More details on the microscopic PRM for $\Lambda$ hypernuclei are
   given in Refs.~\cite{MHYM14,Mei2015}.

\begin{figure}[]
\centering
\includegraphics[width=8cm]{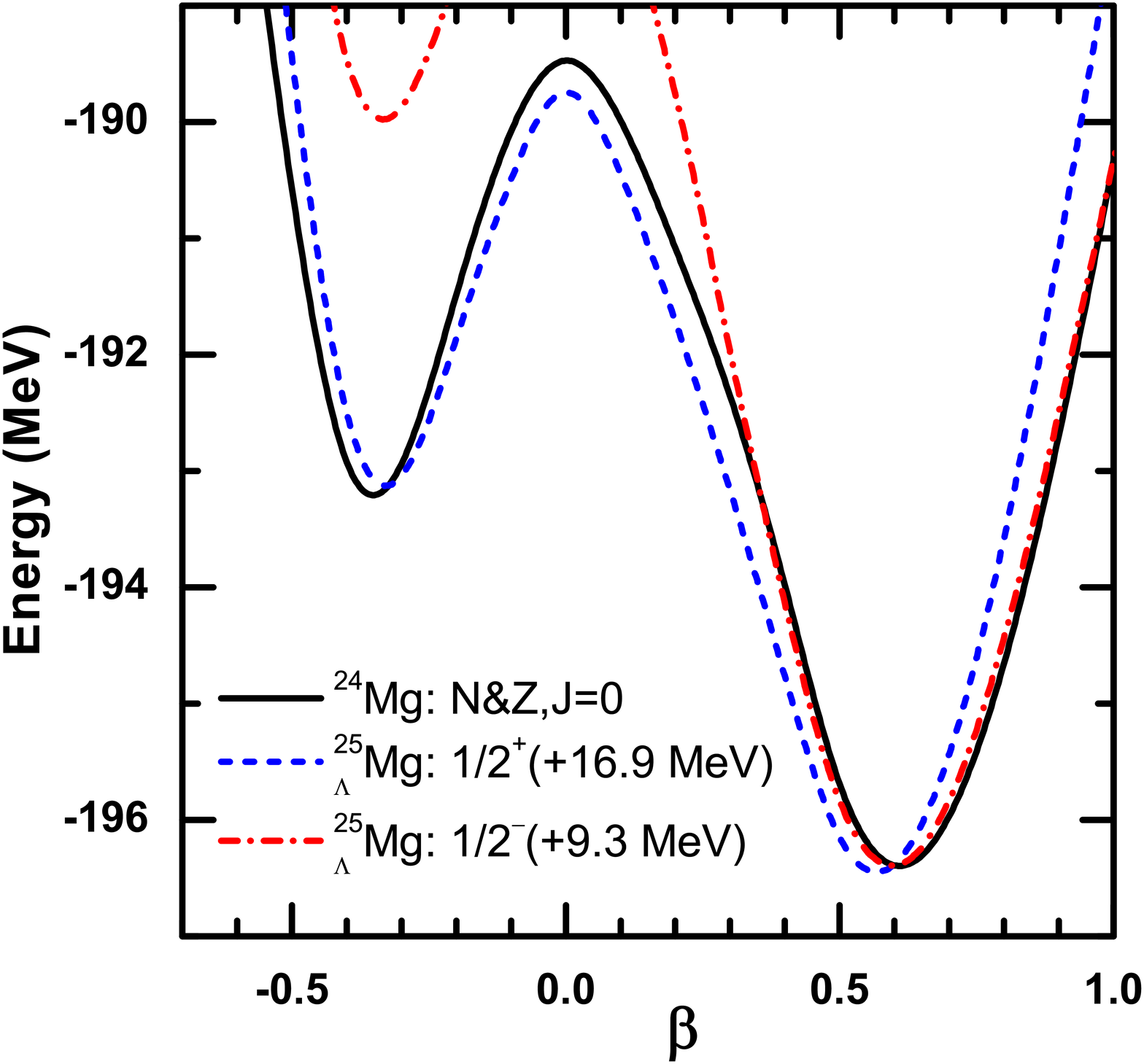}
\caption{(Color online) A comparison of the particle-number and
  angular-momentum projected PES (the solid curve) of $^{24}$Mg
  with the PESs with spin-parity of $I^\pi=1/2^+$ (the dashed curve)
  and $1/2^-$ (the dash-dotted curve) for $^{25}_\Lambda$Mg obtained with
  the microscopic particle-rotor model. }
\label{fig:PRM}
\end{figure}

Figure~\ref{fig:PRM} shows the resultant PES $E_{I}(\beta)$
for $^{25}_{~\Lambda}$Mg with spin-parity of $I^\pi=1/2^+$ and $1/2^-$
as a function of the deformation $\beta$ of the core nucleus. The
energy curve for the $3/2^-$ state is almost the same as the that for
the $1/2^-$ state and is therefore not shown in the figure.
In these calculations, only the leading-order four-fermion coupling
terms (\ref{LO}) are taken into account for the $N\Lambda$ interaction
with coupling strengths fitted to the $\Lambda$ binding
energy from the coupled-channel PRM calculation~\cite{MHYM14} to the value $B_\Lambda=16.6$ MeV from
the triaxial RMF calculation  for $^{25}_{\Lambda s}$Mg.
For comparison, the PES for the core nucleus $^{24}$Mg
with projection onto the particle numbers
and angular momentum $J=0$ is also plotted.
(The PESs for $^{25}_\Lambda$Mg are calculated by coupling the hyperon to
several $J$ states of the core nucleus built on the
deformed mean-field state.)
We note that the $1/2^+$ state
is dominated by the configuration with $\Lambda$ in $s$ orbital,
while the $1/2^-$ state is
dominated by the configurations
of $[p_{1/2}\otimes 0^+]$ and $[p_{3/2}\otimes 2^+]$
at nonzero deformation $\beta$.

The impurity effect of $\Lambda$ in the $s$ and $p$ orbitals
on the PES after restoration of rotational symmetry can be inferred
from the comparison of the PESs for $^{25}_\Lambda$Mg with the
projected PES (N\&Z, $J=0$) for $^{24}$Mg. It is shown that
the PES for $^{25}_\Lambda$Mg with $I^\pi=1/2^+$ has a global energy
minimum at a slightly smaller deformation $\beta=0.55$ than the deformation
$\beta=0.60$ for $^{24}$Mg.  It confirms the conclusion drawn from
the mean-field results shown in Fig.~\ref{fig:Lambdadeform}(a).
For $^{25}_\Lambda$Mg with $1/2^-$, the energies of the spherical
and oblate deformed shapes with respect to the prolate minimum
are significantly increased compared with those for the core nucleus
with $J=0$. However, it is difficult to assess the change of the
collectivity for this case, which depends on the distribution of
the weight function in deformation plane.
In order to determine such distribution,
one could
carry out the microscopic PRM calculation by coupling
the $\Lambda$ to the configuration mixed nuclear core states, as has been
done for $^{9}_\Lambda$Be in Ref. \cite{MHYM14}.
However, this method is currently limited only to axial deformations.
Alternatively, one can introduce a triaxial GCM or 5DCH method
for the core nucleus to examine the $\Lambda$ impurity effect
in a quantitative way. Since the former is very time-consuming
for a systematic study, the 5DCH is adopted in the subsequent study.

 \section{Beyond mean-field study of core nuclei with microscopic collective Hamiltonian method}
 \label{Sec.V}

In the 5DCH approach, the collective excitations of the
core nucleus are described with the following collective Hamiltonian,
 \begin{eqnarray}\label{hamilton}
\hat{H} = \hat{T}_{\textrm{vib}} + \hat{T}_{\textrm{rot}} + V_{\textrm{coll}},
\end{eqnarray}
where the first two terms are the vibrational kinetic energy
\begin{eqnarray}
\hat{T}_{\textnormal{vib}}
 &=&-\frac{\hbar^2}{2\sqrt{wr}}
   \left\{\frac{1}{\beta^4}
   \left[\frac{\partial}{\partial\beta}\sqrt{\frac{r}{w}}\beta^4
   B_{\gamma\gamma} \frac{\partial}{\partial\beta}\right.\right.\nonumber\\
  && \left.\left.- \frac{\partial}{\partial\beta}\sqrt{\frac{r}{w}}\beta^3
   B_{\beta\gamma}\frac{\partial}{\partial\gamma}
   \right]+\frac{1}{\beta\sin{3\gamma}} \left[
   -\frac{\partial}{\partial\gamma} \right.\right.\nonumber\\
  && \left.\left.\sqrt{\frac{r}{w}}\sin{3\gamma}
      B_{\beta \gamma}\frac{\partial}{\partial\beta}
    +\frac{1}{\beta}\frac{\partial}{\partial\gamma} \sqrt{\frac{r}{w}}\sin{3\gamma}
      B_{\beta \beta}\frac{\partial}{\partial\gamma}
   \right]\right\},\nonumber\\
 \end{eqnarray}
and the rotational kinetic energy
\begin{eqnarray}
\hat{T}_{\textrm{rot}} = \frac{1}{2}\sum^3_{k = 1}\frac{\hat{J_k^2}}{{\cal I}_k},
\end{eqnarray}
with $\hat{J_k}$ denoting the components of the angular momentum in the body-fixed frame of a nucleus.
Two quantities that appear in the vibrational kinetic
energy, that is, $r=B_1B_2B_3$ and $w=B_{\beta\beta}B_{\gamma\gamma}-B_{\beta\gamma}^2 $,
determine the volume element in the collective space. The mass parameters
$B_{\beta\beta}$, $B_{\beta\gamma}$ and $B_{\gamma\gamma}$, as well as the
moments of inertia ${\cal I}_k$, depend on the quadrupole
deformation variables $\beta$ and $\gamma$,
\begin{eqnarray}
{\cal I}_k = 4B_k \beta^2\sin^2(\gamma - 2k \pi/3), k = 1, 2, 3,
\end{eqnarray}
and are determined by the triaxial RMF+BCS calculations in cranking approximation~\cite{Niksic2009PhysRevC.79.034303}.

The third term
$V_{\textrm{coll}}$ in Eq. (\ref{hamilton}) is a collective potential given by
\begin{equation}\label{Vcoll}
V_\textrm{coll}(\beta, \gamma)=\tilde E_{\rm tot} -\triangle V_{\textrm{vib}}(\beta, \gamma)-\triangle V_{\textrm{rot}}(\beta, \gamma),
\end{equation}
where $\triangle V_{\textrm{vib}}(\beta, \gamma)$
and $\triangle V_{\textrm{rot}}(\beta, \gamma)$ are the zero-point-energy
of vibrational and rotational motions, respectively.
The $\tilde E_{\rm tot}$ is given by the summation of total energy
for the  core nucleus inside the hypernucleus and the  $N\Lambda$
interaction energy $\varepsilon_{N\Lambda}(\br)$ (\ref{Energy:NL})
which carries most of information on the $\Lambda$ impurity effect.

The eigenvalue problem with the collective Hamiltonian is solved by expanding
eigenfunctions in terms of a complete set of basis functions
that depend on the five degrees of freedom: the deformation
variables $\beta$ and $\gamma$, and the Euler
angles $\phi$, $\theta$ and $\psi$~\cite{Prochniak1999_NPA648-181}.

\begin{figure}
\centering
\includegraphics[width=9cm]{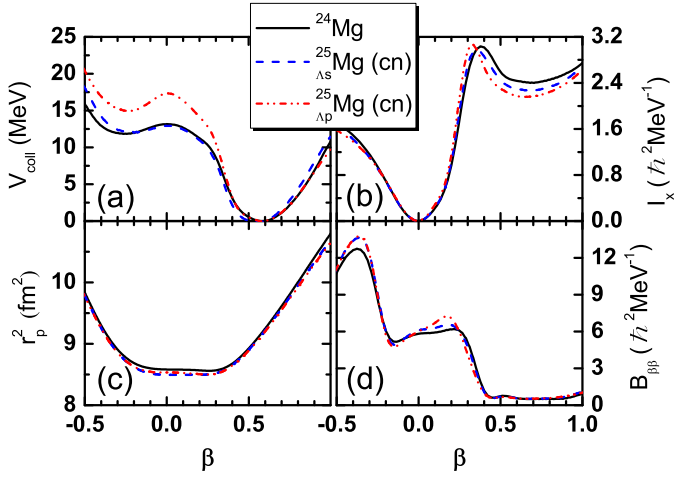}
\caption{(Color online) (a) The collective potential $V_{\rm coll}$,
  (b) the moment of inertia along $x$-axis ${\cal I}_x$, (c) the mean
  squared radius of protons, and (d) the mass parameters $B_{\beta\beta}$
  as a function of quadrupole deformation $\beta$ for
  $^{24}$Mg and $^{25}_{\Lambda}$Mg (cn) obtained with the fine-dimensional
  collective Hamiltonian (5DCH) approach.}
\label{fig:collective24}
\end{figure}

\begin{figure}[]
\centering
\includegraphics[width=8cm]{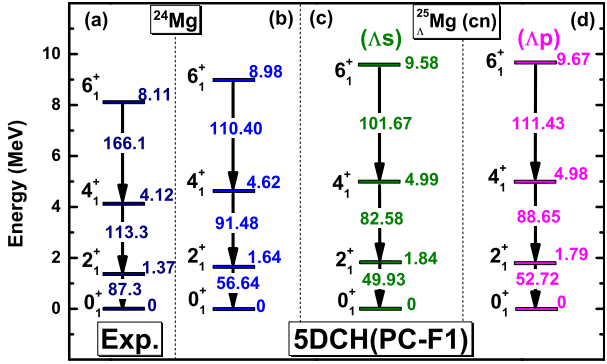}
\caption{(Color online) The low-spin spectra of
  the ground state band for $^{24}$Mg (b) and
  the nuclear core of $^{25}_{~\Lambda}$Mg (c, d) obtained with the 5DCH method.
  The $B(E2)$ values are in units of e$^2$fm$^4$. The spectrum of $^{24}$Mg
  is compared with the corresponding experimental data, taken
  from Ref. \cite{Endt1990_NPA510-1}.}
\label{fig:Spec24Mg}
\end{figure}

Figure~\ref{fig:collective24} shows the parameters in
the 5DCH as a function of quadrupole deformation $\beta$ for
$^{24}$Mg and the core nucleus inside $^{25}_{\Lambda s}$Mg and $^{25}_{\Lambda p}$Mg.
These are the collective potential $V_{\rm coll}$, the moment of inertia
along $x$-direction ${\cal I}_x$, the rms radius of protons,
and the mass parameters $B_{\beta\beta}$.
It is shown that the collective potentials for the core nucleus
inside $^{25}_{\Lambda s}$Mg and $^{25}_{\Lambda p}$Mg have a similar behavior
as the projected PESs for  $^{25}_{\Lambda s}$Mg and $^{25}_{\Lambda p}$Mg
shown in Fig.~\ref{fig:PRM}.
Moreover, the moment of inertia for the core nucleus around the
energy minimum is significantly reduced by $\Lambda_s$ and $\Lambda_p$.
The resultant energy spectrum is stretched as shown in Fig.~\ref{fig:Spec24Mg}
for the low-spin spectra of ground state band for the $^{24}$Mg
and the core nucleus inside $^{25}_{\Lambda}$Mg. The $\Lambda_s$
increases the excitation energy $E_x(2^+_1)$ for the
$2_1^+$ state by $\sim12.2\%$ and reduces the $E2$ transition
strength $B(E2: 2_1^+\rightarrow 0_1^+$) by $\sim$ 11.8$\%$, compared
with the value of $\sim7\%$ and $\sim9\%$, respectively,
found in our previous 5DCH calculation for the same nucleus
based on the non-relativistic Skyrme EDFs~\cite{Yao2011_NPA868-12,Mei2012}.

\begin{figure}[t]
\includegraphics[width=9cm]{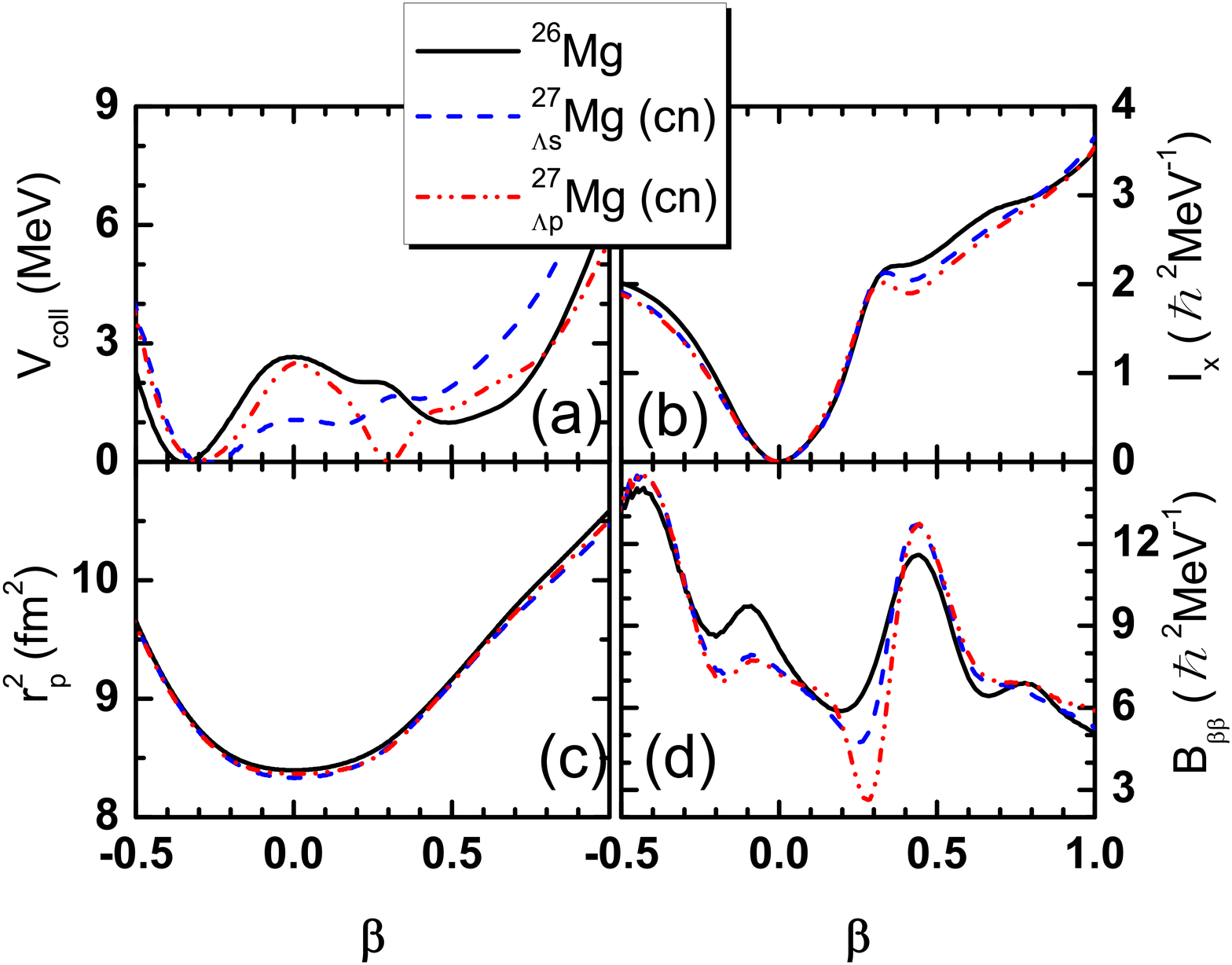} 
\caption{(Color online) Same as Fig. \ref{fig:collective24}, but for $^{26}$Mg and nuclear core inside $^{27}_{\Lambda}$Mg.}
\label{fig:collective26}
\end{figure}
\begin{figure}[]
\includegraphics[width=8cm]{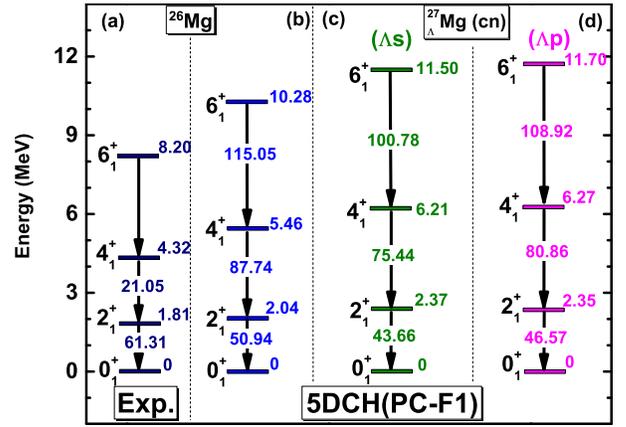}
\caption{(Color online) Same as Fig. \ref{fig:Spec24Mg},
  but for $^{26}$Mg and $_\Lambda^{27}$Mg (cn). The experimental
  data for $^{26}$Mg are taken from Refs. \cite{lbl,nndc}.}
\label{fig:SPE26Mgsp}
\end{figure}

The sensitivity of $\Lambda$ impurity effect on nuclear collective
properties to the underlying EDFs has been examined  based on several
sets of Skyrme EDFs with various pairing strengths
for $^{45}_{\Lambda s}$S~\cite{Mei2012}. It has been found that
although different Skyrme EDFs give somewhat different low-lying
spectra for the core nucleus, they give similar and generally small
size of $\Lambda$ impurity effect (typically within 5\%) on the
spectroscopic observables. From this point of view,
one can draw a conclusion that the present relativistic study
yields the $\Lambda_s$ impurity effect on nuclear low-energy
structure which is larger than that by non-relativistic Skyrme EDFs, similarly to
the conclusion for mean-field calculations~\cite{Schulze2010PTP}.
In addition, it is shown that the $\Lambda_p$
increases the excitation energy of $2_1^+$ state by $\sim$ 9.1$\%$
and reduces the $E2$ transition strength
$B(E2 : 2_1^+\rightarrow 0_1^+)$ by $\sim$ 6.9$\%$.
Moreover, we note that the excitation energy of $2^+_2$ state
is increased from 5.62 MeV to 5.88 MeV and 6.67 MeV
due to the $\Lambda_s$ and $\Lambda_p$, respectively, which is
consistent with the observation in Fig.~\ref{fig:gamma}.

\begin{figure}[t]
\includegraphics[width=9cm]{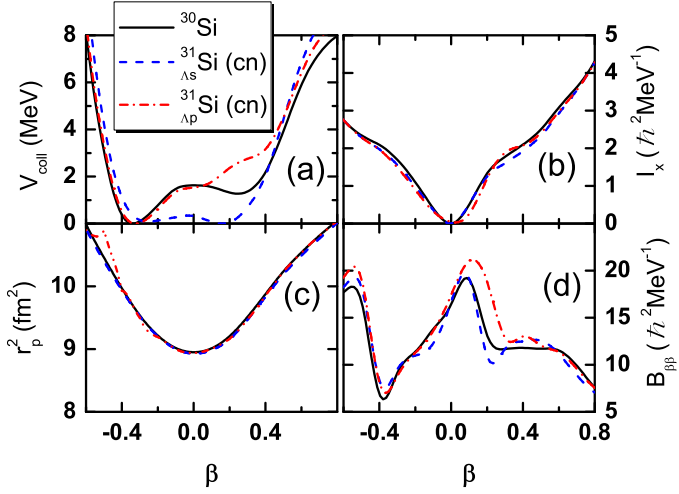}
\caption{(Color online) Same as Fig. \ref{fig:collective24}, but for $^{30}$Si and nuclear core inside $^{31}_{\Lambda}$Si.}
\label{fig:collective30}
\end{figure}
\begin{figure}[]
\includegraphics[width=8cm]{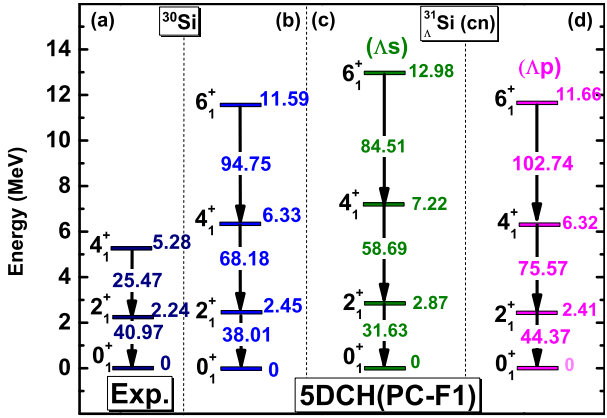}
\caption{(Color online) Same as Fig. \ref{fig:Spec24Mg}, but for $^{30}$Si and $_\Lambda^{31}$Si (cn).
  The experimental data for $^{30}$Si are taken from Refs. \cite{lbl,nndc}.}
\label{fig:SPE30Sisp}
\end{figure}
\begin{figure}[]
\centering
\includegraphics[width=7cm]{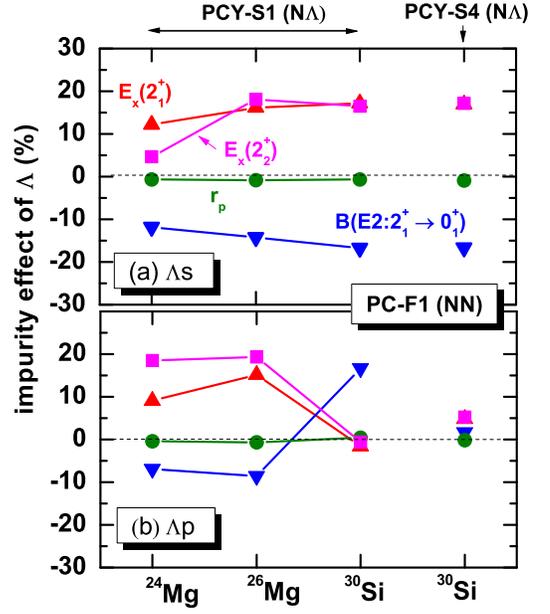}
\caption{(Color online) The impurity effect of $\Lambda_s$ and $\Lambda_p$ hyperon on the excitation energies $E_x(2^+_1)$ and $E_x(2^+_2)$ for the lowest two $2^+$ states, the proton root-mean-squared radius $r_p$ of ground state ($0^+_1$), and the $E2$ transition strength $B(E2: 2_1^+\rightarrow 0_1^+)$ in $^{24,26}$Mg and $^{30}$Si from the 5DCH calculations based on the PC-F1 and PCY-S1 forces for the $NN$ and $N\Lambda$ interactions, respectively. For comparison, the impurity effect of $\Lambda_s$ and $\Lambda_p$ on $^{30}$Si calculated with the $N\Lambda$ interaction of PCY-S4~\cite{Tanimura2012_PRC85-014306} is also plotted in the last column.}
\label{fig:ratio}
\end{figure}

Figure~\ref{fig:collective26} shows the parameters in the 5DCH
for $^{26}$Mg and the core nuclei inside $^{27}_{\Lambda s}$Mg and
$^{26}_{\Lambda p}$Mg. Similar to the PESs in Fig.~\ref{fig:PES26Mg},
the collective potential $V_{\rm coll}$ is rather different for
the $^{27}_{\Lambda s}$Mg, $^{27}_{\Lambda p}$Mg and the core nucleus $^{26}$Mg.
Similarly to the $^{24}$Mg case, the $\Lambda_s$ and $\Lambda_p$ apparently reduce the moments of inertia around the prolate minimum of the collective potential.
The competition of the changes in the collective potential and the
collective parameters results in the $\Lambda$ impurity effect on
the low-energy excitations. Figure~\ref{fig:SPE26Mgsp} shows
the low-spin spectra of the ground-state band for the $^{26}$Mg
and $_\Lambda^{27}$Mg.  It is shown  that the $\Lambda_s$
increases the excitation energy of the $2_1^+$ state by $\sim$ 16.2$\%$
and reduces the $B(E2: 2_1^+\rightarrow 0_1^+)$ by $\sim$ 14.3$\%$.
Moreover, the $\Lambda_p$ increases the excitation energy
of the $2_1^+$ state by $\sim$ 15.2$\%$ and reduces
the $B(E2 : 2_1^+\rightarrow 0_1^+)$ by $\sim$ 8.6$\%$. The
excitation energy of the $2^+_2$ state is increased from 3.92 MeV
to 4.63 MeV and 4.68 MeV due to the $\Lambda_s$ and $\Lambda_p$, respectively.

Figure~\ref{fig:collective30} shows the collective parameters
for $^{30}$Si and $_\Lambda^{31}$Si. In contrast to the $\Lambda_p$ effect
in $^{25,27}_{\Lambda s}$Mg, the $\Lambda_p$ increases the mass parameters
in most of the deformation regions. Together with the well-developed
oblate minimum, the $\Lambda_p$ increases the nuclear collectivity
as shown in Fig.~\ref{fig:SPE30Sisp}. The $\Lambda_p$ increases
significantly the $B(E2: 2_1^+\rightarrow 0_1^+)$ of the
core nucleus $^{30}$Si by $\sim 16.7\%$ and decreases
slightly the excitation energy of $2_1^+$ state. The excitation
energy of the $2^+_2$ state is altered
from 4.28 MeV to 4.99 MeV and 4.26 MeV due to the $\Lambda_s$
and $\Lambda_p$,  respectively.

The impurity effect of $\Lambda_s$ and $\Lambda_p$ is summarized
in Fig.~\ref{fig:ratio}. The change in the proton root-mean-squared radius of ground state
is within 1\% for all the three nuclei. It confirms that the
reduction/enhancement of the $B(E2)$ value for the core nucleus by
adding a $\Lambda$ particle mainly originates from the modification of
nuclear collective potential, the moment of inertia, and the
mass parameters, rather than a shrinkage or an expansion of
proton distribution as found in light nuclear systems
with cluster structures~\cite{HKMM99}. One can also see that the changes
in the excitation energies $E_x(2^+_1)$ and $E_x(2^+_2)$ for the lowest two
$2^+$ states are similar to one another. To examine the model-dependence of the $\Lambda$ impurity effect for $^{30}$Si, we also plot the results by the PCY-S4 $N\Lambda$ interaction in Fig.~\ref{fig:ratio}. It shows again that the impurity effect of $\Lambda_s$ is much less sensitive to the $N\Lambda$ interaction than that of $\Lambda_p$,
that is consistent with the potential energy surfaces in Fig.~\ref{fig:30Si-comp}.

%
\section{summary}\label{Sec.VI}
%
%

We have established a triaxially deformed relativistic
mean-field approach for $\Lambda$ hypernuclei based on
a point-coupling EDF. Using the 5DCH method based on this approach, we have
quantitatively studied the impurity effect of $\Lambda_s$ and $\Lambda_p$
hyperon on the low-energy collective excitations of $^{24}$Mg, $^{26}$Mg,
and  $^{30}$Si. Besides, the quadrupole deformation effect on the $\Lambda$ binding
energies of hypernuclei has been studied. In particular, the PESs of
three $sd$-shell $\Lambda$ hypernuclei $^{25,27}_{~~~~\Lambda}$Mg and $^{31}_{~\Lambda}$Si,
as well as their core nucleus in $(\beta, \gamma)$ deformation plane has been calculated.
The low-lying states of the core nuclei before and after adding $\Lambda$ hyperon in the lowest
positive-parity ($\Lambda_s$) and the negative-parity ($\Lambda_p$)
states have also been discussed. Moreover, the PESs of $^{25}_{\Lambda}$Mg with spin-parity of
$I^\pi=1/2^+$ and $1/2^-$ have been obtained with the microscopic PRM and compared with the PES of the
core nucleus with $J^\pi=0^+$. Our main findings in the present studies are summarized as follows:

\begin{itemize}
\item The quadrupole deformation decreases the $\Lambda_s$ binding
  energy and increases the $\Lambda_p$ binding energy in $\Lambda$ hypernucleus.

\item The potential energy surfaces of the whole $\Lambda$ hypernuclei
  could be significantly different from those of the core nuclei without
  the hyperon impurity.  In general, the hypernuclei with a  $\Lambda_s$
  ($\Lambda_p$) have an energy minimum with smaller (or larger) deformation
  than the core nucleus.  However, the potential energy surfaces of
  the core nuclei inside the hypernuclei are very similar to that of
  the nuclei without hyperon.

\item Quantitatively, the $\Lambda_s$ increases the excitation energy
  of the $2^+_1$ state and decreases the $E2$ transition strength from
  this state to the ground state in the core nucleus by $12\%-17\%$,
  about twice larger than the value found in our previous 5DCH study
  based on the non-relativistic Skyrme EDFs.
  However, $\Lambda_p$ can either increase or decrease the
  collectivity of the core nucleus depending on the competition
  between the changes in the potential energy surface and the
  collective parameters.
\end{itemize}

Finally, we emphasize that the generalization of our triaxial RMF approach for hypernuclei to multi-strangeness systems is straightforward.  Moreover, the present approach provides a starting point to carry out a beyond mean-field calculation of the low-lying states of $\Lambda$ hypernuclei by introducing the techniques of exact projections and GCM~\cite{Yao10,Yao14} for the odd-mass system, the results of which can be compared with those of the microscopic PRM~\cite{MHYM14} based on the same relativistic point-coupling EDF.

\begin{acknowledgements}
  We thank E. Hiyama, T. Koike, M. Isaka and S. G. Zhou for fruitful
  discussions. W. X. X. also thanks the Institute of Theoretical Physics
  in Chinese Academy of Sciences and the Nuclear Theory Group in Tohoku
  University for their hospitality during her visit. This work was supported
  by the NSFC under Grant Nos. 11475140, 11305134, 11105110, and 11105111,
  the Tohoku University Focused Research Project ``Understanding
  the origins for matters in universe" and JSPS KAKENHI Grant Number 26400263.
\end{acknowledgements}

\bibliographystyle{apsrev4-1}
%

\end{document}